\documentclass[11pt,a4paper]{article}
\usepackage{jheppub}
\DeclareMathOperator{\tr}{tr} 
\DeclareMathOperator{\Spin}{Spin} 

\title{Ground States of Duality-twisted Sigma-Models with $K_3$ Target Space}
\author{Ori J. Ganor, Sharon Jue, and Shannon McCurdy}
\affiliation{
Department of Physics,
  University of California,\\
Berkeley, CA 94720, U.S.A.}
\emailAdd{origa@socrates.berkeley.edu}
\emailAdd{sljue@berkeley.edu}
\emailAdd{smccurdy@berkeley.edu}

\abstract{
We analyze the ground states of a two-dimensional sigma-model whose target space is an elliptically fibered $K_3$, with the sigma-model compactified on $S^1$ with boundary conditions twisted by a duality symmetry. We show that the Witten index receives contributions from two kinds of states: (i) those that can be mapped to cohomology with coefficients in a certain line bundle over the target space, and (ii) states whose wave-functions are localized at singular fibers. We also discuss the orbifold limit and possible connections with geometric quantization of the target space.
}

\begin{document}
\maketitle
\flushbottom

\newcommand{\secref}[1]{\S\ref{#1}}
\newcommand{\figref}[1]{Figure~\ref{#1}}
\newcommand{\appref}[1]{Appendix~\ref{#1}}
\newcommand{\tabref}[1]{Table~\ref{#1}}

\newcommand\SUSY[1]{{${\mathcal{N}}={#1}$}}  
\newcommand\px[1]{{\partial_{#1}}}

\def\be{\begin{equation}}
\def\ee{\end{equation}}
\def\bear{\begin{eqnarray}}
\def\eear{\end{eqnarray}}
\def\nn{\nonumber}

\newcommand\bra[1]{{\left\langle{#1}\right\rvert}} 
\newcommand\ket[1]{{\left\lvert{#1}\right\rangle}} 

\newcommand{\C}{\mathbb{C}}
\newcommand{\R}{\mathbb{R}}
\newcommand{\Z}{\mathbb{Z}}
\newcommand{\CP}{\mathbb{CP}}

\def\SL{{{\mbox{\rm SL}}}} 

\def\xa{{\mathbf{a}}} 
\def\xb{{\mathbf{b}}} 
\def\xc{{\mathbf{c}}} 
\def\xd{{\mathbf{d}}} 

\def\TGT{{X}}
\def\OpM{{\hat{\mathfrak{M}}}} 
\def\Op{{\hat{\mathcal{O}}}} 
\def\Ham{{\hat{H}}} 

\def\sz{{z}}
\def\bsz{{\overline{\sz}}}
\def\bpartial{{\overline{\partial}}}
\def\smX{{\phi}} 

\def\bz{{\overline{z}}}
\def\zF{{\mathfrak{w}}} 
\def\bzF{{\overline{\zF}}} 
\def\zB{{\mathfrak{z}}} 
\def\bzB{{\overline{\zB}}} 

\def\oi{{\overline{i}}}
\def\oj{{\overline{j}}}
\def\ok{{\overline{k}}}
\def\om{{\overline{m}}}

\def\bpsi{{\overline{\psi}}}

\def\symPW{{\mathfrak U}} 

\def\zg{{\mathbf{g}}} 
\def\bQ{{\overline{Q}}} 
\def\a{\alpha}
\def\b{\beta}
\def\dta{{\dot{\a}}}
\def\dtb{{\dot{\b}}}

\def\Hilb{{\mathcal{H}}}
\def\Hgq{{\mathcal{H}_{\text{gq}}}} 

\def\Lbd{{\mathcal{L}}} 
\def\Kah{{K}} 
\def\Vbd{{\mathcal{V}}}

\def\slt{{(<)}} 
\def\sgt{{(>)}} 

\def\phiMode{{\alpha}} 
\def\wphiMode{{\tilde{\alpha}}} 

\def\psiMode{{\mathfrak{b}}} 
\def\bpsiMode{{\overline{\psiMode}}}  

\def\ttX{{x}}

\def\pSUSY{{\eta}}
\def\bpSUSY{{\overline{\eta}}}
\def\cD{{\mathcal D}}

\def\xL{{L}} 

\def\dimTGT{{d}} 

\def\efBase{{\mathfrak{B}}} 
\def\efFiber{{\mathfrak{F}}} 
\def\xf{{\mathbf{x}}} 
\def\yf{{\mathbf{y}}} 
\def\xyf{{\mathbf{z}}} 
\def\lvk{{\mathbf{k}}}
\def\Lag{{L}} 
\def\BerryA{{{\mathcal A}}} 
\def\BerryD{{{\mathcal D}}} 
\def\Obd{{{\mathcal O}}} 

\def\ModSpCP{{\mathcal{M}}} 
\def\fTh{{\varphi}} 
\def\Ttau{{\mathfrak{E}}} 
\def\tLbd{{\widetilde{\Lbd}}} 
\def\BerryActr{{\mathfrak a}} 
\def\LoopC{{\mathcal C}} 
\def\btau{{\overline{\tau}}}
\def\vertBF{{\Upsilon}} 
\def\mvS{{\mathcal S}} 
\def\mvT{{\mathcal T}} 
\def\mvR{{\mathcal R}} 
\def\mvM{{\OpM}} 

\def\scX{{X}} 
\def\wpsi{{\tilde{\psi}}} 
\def\Zivp{{\Upsilon}} 
\def\Zivw{{\Omega}} 

\def\wtau{{\widetilde{\tau}}} 
\def\wxf{{\widetilde{\xf}}} 

\def\Id{{\mathbb{I}}} 

\def\wfB{{\lambda}} 
\def\wfF{{\varphi}} 

\def\NCzeta{{\widetilde{\zeta}}} 

\def\brho{{\overline{\rho}}}
\def\ord{{\mathbf{r}}} 

\def\One{{\mathfrak 1}}
\def\bOne{{\overline{\One}}}
\def\Two{{\mathfrak 2}}
\def\bTwo{{\overline{\Two}}}
\def\zi{{\mathfrak z}}
\def\bzi{{\overline{\zi}}}
\def\wi{{\mathfrak w}}
\def\bwi{{\overline{\wi}}}

\def\zS{{\mathfrak z}} 
\def\bzS{{\overline{\zS}}} 

\def\losc{{\alpha}} 
\def\rosc{{\widetilde{\alpha}}} 

\def\jQ{{\mathcal J}} 

\def\Mf{{\mathcal M}} 
\def\Curve{{\mathcal C}} 

\def\tPhi{{\widetilde{\Phi}}} 
\def\canMom{{\Pi}} 

\def\omegaF{{\omega}} 

\def\Patch{{\mathcal U}} 
\def\CanBdl{{\mathcal K}} 
\def\FD{{\mathcal F}} 

\def\osc{{\alpha}} 
\def\wosc{{\tilde{\alpha}}} 

\def\MHitchin{{{\mathcal M}_H}} 



\section{Introduction}
\label{sec:Intro}

Consider a supersymmetric 1+1D $\sigma$-model with target space $\TGT$ (a K\"ahler manifold with metric $G$ and $2$-form field $B$) that is invariant under a suitable mirror-symmetry transformation, so that in this model mirror-symmetry is a discrete nonperturbative symmetry. We wish to study the Hilbert space $\Hilb$ of ground states of a compactification of this $\sigma$-model on $S^1$ with supersymmetric boundary conditions that include a {\it mirror-symmetry twist}. The mirror-symmetry twist is constructed as follows. Let $(\sigma^0,\sigma^1)$ be the space-time coordinates with $0\le\sigma^1<2\pi$ parameterizing $S^1$, and $-\infty<\sigma^0<\infty$ parameterizing time. To insert a mirror-symmetry twist at, say, $\sigma^1=0$, we first Wick rotate by setting $\sigma^0=i\sigma^2$, then consider $\sigma^1$ as the (Euclidean) time direction and insert the operator $\OpM$ that realizes mirror-symmetry at $\sigma^1=0.$ Correlators of operators $\Op_1,\dots,\Op_n$ in this theory can be defined as $\tr\{\OpM(-1)^F e^{-2\pi\Ham}\Op_n\cdots\Op_1\}$, where $\Ham$ is the Hamiltonian of the field theory, $F$ is the fermion number, $\tr$ is the trace over the Hilbert space of the (untwisted) $\sigma$-model compactified on $S^1$ (which now corresponds to the space direction $\sigma^2$), and the operators are assumed to be in the Heisenberg picture and ordered according to increasing $\sigma^1$.

Duality twists (sometimes referred to as ``monodrofolds'' or, in our context, as ``mirrorfolds'') have been extensively studied in a variety of contexts (see for example \cite{Kumar:1996zx}-\cite{Kawai:2007nb} for a sampling of the literature), and several interesting approaches have been developed to understand such backgrounds. For $T^{2d}$ target spaces and their orbifolds the mirror-twist reduces to a T-duality twist, and the compactification is a sector of an asymmetric orbifold. For a $K_3$ target space, the spectrum of the theory at special limit points in moduli space (Gepner points) was computed in \cite{Kawai:2007nb}.

Our goal in this paper is relatively modest -- to learn about the ground states of the theory, describe their wave-functions in more detail, and explore possible connections with geometric quantization of the target space.
One of the motivations for this work is to develop tools to study compactifications with duality twists in gauge theory. For example, a compactification of 3+1D \SUSY{4} Super-Yang-Mills theory on $S^1$ with an S-duality twist was studied in \cite{Ganor:2008hd}-\cite{Ganor:2012mu}, where a possible connection with Chern-Simons theory was also explored. This hints at a relation between our present $\sigma$-model problem and geometric quantization as follows. S-duality can be related to mirror-symmetry of a certain 1+1D $\sigma$-model by compactifying the 3+1D gauge theory on a Riemann surface \cite{Harvey:1995tg,Bershadsky:1995vm}. The $\sigma$-model's target space is the {\it Hitchin space} associated with the gauge group and the Riemann surface (see \cite{Kapustin:2006pk} for more details). On the other hand, the Hilbert space of Chern-Simons theory on a Riemann surface can be obtained by geometric quantization of the moduli space of flat connections on the surface \cite{Witten:1988hf}, and the latter can naturally be embedded in the Hitchin space as the fixed locus of a certain global symmetry. Thus, the question arises whether the Hilbert space of the mirror-symmetry twisted $\sigma$-model compactification is related to the Hilbert space obtained by geometric quantization of the target space (or a suitable subspace of it).

Another motivation is to learn about the operator $\OpM$ that realizes mirror-symmetry.
Mirror-symmetry is a nonperturbative phenomenon of 1+1D $\sigma$-models, and the Hilbert space of the mirror-twisted compactification carries information about the operator $\OpM.$ Restricting to ground states is equivalent to taking the low-energy limit, where roughly speaking, correlators reduce to $\tr\{\OpM(-1)^F \Op_n\cdots\Op_1\}.$ Thus, understanding the Hilbert space of ground states (and the operators that act on it) is a way to probe the mirror symmetry operator $\OpM$.

The main focus of this paper is one particular case (with potential generalizations) --- an elliptically fibered $K_3$ target space, with the mirror-symmetry that can be realized as T-duality on the fiber.
While a lot of progress has been made in understanding mirror-symmetry (see e.g., \cite{Hori:2000kt,Hori:2003ic} and references therein)
and certainly a lot is known about mirror-symmetry of $K_3$, we are unaware of any explicit study of the ground states of the particular compactification that we are investigating in this paper.
For the present analysis we begin with a careful study of a $\sigma$-model with $T^2$ target space, compactified with a T-duality twist. The Hilbert space of ground states naturally maps to the Hilbert space obtained by deformation quantization of $T^2$ at level $2$. We then fiber this Hilbert space over the base of the elliptic fibration to construct ground states of the $K_3$ compactification. We then calculate the contribution of these states to the Witten index (which can be calculated by other means), and we find a discrepancy. We interpret this discrepancy as the contribution of ground states with wave-functions localized at singular fibers. We show that the two calculations of the Witten index are consistent if one ground state is associated with each singular fiber, and we bring supporting evidence for this conjecture from nonperturbative string theory.

The paper is organized as follows.
In \secref{sec:MirrorTwist} we review some basic facts about supersymmetric nonlinear $\sigma$-models and define the mirror-twist.
In \secref{sec:T2} we discuss the simple case of a $T^2$ target space.
In \secref{sec:EFK3} we discuss the case of elliptically fibered $K_3$ surfaces with a mirror-symmetry twist that can be understood as local T-duality of the fiber. This section contains our main results for the ground states.
In \secref{sec:OtherX} we discuss duality twists in the orbifold limit of the $K_3$ $\sigma$-model, and in \secref{sec:ConnectGQ} we further explore possible connections with geometric quantization.
We conclude with a discussion in \secref{sec:disc}.

\section{The $\sigma$-model and the mirror-symmetry twist}
\label{sec:MirrorTwist}

Our spacetime is two-dimensional, and we work either in Minkowski signature with spacetime coordinates $(\sigma^0,\sigma^1)$ or in Euclidean signature with complex (worldsheet) coordinate $\sz\equiv\sigma^1+i\sigma^2$.
The supersymmetric $\sigma$-models have an action of the form \cite{Zumino:1979et} (and we follow the conventions of \cite{Witten:1991zz}):
\bear
I_0 &=&\tfrac{1}{\pi} \int\bigl(
\tfrac{1}{2}(G_{IJ}+i B_{IJ})\partial\smX^I\,\bpartial\smX^J
+\tfrac{i}{2}G_{IJ}\psi^I\partial\psi^J
+\tfrac{i}{2}G_{IJ}\bpsi^I\bpartial\,\bpsi^J
\nn\\ &&
+\tfrac{i}{2}G_{IJ}\psi^I\partial\smX^L\Gamma^J_{LK}\psi^K
+\tfrac{i}{2}G_{IJ}\bpsi^I\bpartial\smX^L\Gamma^J_{LK}\bpsi^K
+\tfrac{1}{4}R_{IJKL}\psi^I\psi^J\bpsi^K\bpsi^L
\bigr)d^2\sz\,,
\label{eqn:SigmaModelI}
\eear
where $\partial\equiv\partial_\sz$, $\bpartial\equiv\partial_\bsz$,
$\smX^I$ ($I=1,\dots,d$) are scalars, $\psi^I$ and $\bpsi^I$ are fermions,
$G_{IJ}(\smX)$ are the components of the metric on target space $\TGT$, $\Gamma^J_{LK}$ are the Christoffel symbols derived from $G_{IJ}$,
$R_{IJKL}$ are the components of curvature, and $B_{IJ}(\smX)$ are the components of an antisymmetric $B$-field, assumed to be a closed $2$-form.
We will take $\TGT$ to be a Calabi-Yau manifold and label the complex coordinates by $i,j,k,\dots$ and their complex conjugates by $\oi,\oj,\ok,\ldots$. The K\"ahler condition is equivalent to requiring that the metric $G_{IJ}$ is of type $(1,1)$ (i.e., $G_{i\oi}=G_{\oi i}$ are the only possible nonzero components) and further, the only allowed nonzero Christoffel symbols are $\Gamma^i_{jk}$ and $\Gamma^\oi_{\oj\ok}$. The Calabi-Yau condition is equivalent to having a coordinate system where $\sqrt{G}=\det G_{i\oi}=1$.
We also assume that the $2$-form $B$-field is of type $(1,1)$ (i.e., only $B_{i\oi}=-B_{\oi i}$ can be nonzero).

Next, we introduce the mirror-symmetry-twist. We assume that $\TGT$ and the metric $G$ and $B$-field are special so that ``mirror-symmetry'' is not only a duality but an actual symmetry. In other words, the mirror of $\TGT$ is the same manifold and at the same point in moduli-space. We then introduce a cut at $\sigma^1=0$ (for all times $\sigma^0$) and connect the fields just left of the cut to the fields just right of the cut via a mirror-symmetry transformation (which, of course, is not algebraic in the fields). Our goal is to understand the low-energy description of this system, i.e., the ground states.

The Lagrangian \eqref{eqn:SigmaModelI} is invariant under the supersymmetry transformations:
\bear
\delta\smX^I &=& i\pSUSY\psi^I+i\bpSUSY\bpsi^I\,,\nn\\
\delta\psi^I &=& 
-\pSUSY\partial\smX^I-i\bpSUSY\bpsi^K\Gamma^I_{KM}\psi^M\,,
\label{eqn:SusyI}\\
\delta\bpsi^I &=& 
-\bpSUSY\bpartial\smX^I-i\pSUSY\psi^K\Gamma^I_{KM}\bpsi^M\,,
\nn
\eear
where $\pSUSY$ and $\bpSUSY$ are left and right SUSY generators. 
Mirror-symmetry commutes with these transformations.

For a K\"ahler target space, we split the coordinates into holomorphic (labeled by $i,j,k,\dots$) and anti-holomorphic ($\oi,\oj,\ok,\dots$).
The nonzero metric components are of $(1,1)$-type (i.e., $G_{i\oj}$) and satisfy the K\"ahler condition ($\partial_{[i} g_{j]\ok}=0$).
Defining $\bpsi_i\equiv G_{i\oj}\bpsi^\oj$ and $\psi_\oi\equiv G_{j\oi}\psi^j$, we can write the action as
\bear
I_0 &=&
\tfrac{1}{\pi} 
\int\bigl(
G_{i\oj}\partial\smX^i\,\bpartial\smX^\oj
+i \psi_\oj\bpartial\psi^\oj
+i \bpsi_i\partial\,\bpsi^i
\nn\\ &&
+i\bpartial\smX^\oi\Gamma^\oj_{\oi\ok}\psi_\oj\psi^\ok
+i\partial\smX^l\Gamma^i_{lk}\bpsi_i\bpsi^k
+{{{R^\ok}_\oj}^i}_j\bpsi_i\bpsi^j\psi_\ok\psi^\oj
\bigr)d^2\sz
\label{eqn:SigmaModelKII}
\eear

For every complex structure under which the target space is K\"ahler there are additional SUSY transformations:
\bear
\delta\smX^i &=& 0\,,
\quad
\delta\smX^\oi = i\pSUSY\psi^\oi+i\bpSUSY\bpsi^\oi\,,
\nn\\
\delta\psi^i &=& -\pSUSY\partial\smX^i\,,
\quad
\delta\psi^\oi = -i\bpSUSY\bpsi^\oj\Gamma^\oi_{\oj\ok}\psi^\ok\,,
\label{eqn:SusyII}\\
\delta\bpsi^i &=& -\bpSUSY\bpartial\smX^i\,,
\quad
\delta\bpsi^\oi = -i\bpSUSY\psi^\oj\Gamma^\oi_{\oj\ok}\bpsi^\ok\,.
\nn
\eear
Mirror-symmetry does not necessarily commute with these transformations.
We will return to the question of how much SUSY is preserved in \secref{subsec:SUSY}.

Later on, it will be convenient to have explicit expressions for the canonical duals of $\smX^i$ and $\smX^\oi$. They are:
$$
\Pi_i = \tfrac{1}{4}G_{i\oj}\partial_0\smX^\oj
+\tfrac{1}{2}i\Gamma^k_{ij}\bpsi_k\bpsi^j
\,,\qquad
\Pi_\oi = \tfrac{1}{4}G_{j\oi}\partial_0\smX^j
+\tfrac{1}{2}i\Gamma^\ok_{\oi\oj}\psi_\ok\psi^\oj
\,.
$$
and we have
$$
[\Pi_i(\sigma_1),\smX^j(\sigma_1')]= -i\delta_i^j\delta(\sigma_1-\sigma_1')\,,\qquad
[\Pi_\oi(\sigma_1),\smX^\oj(\sigma_1')]= -i\delta_\oi^\oj\delta(\sigma_1-\sigma_1')\,.
$$
and the fermions' commutation relations are
$$
\{\psi^\oi(\sigma_1),\psi_\oj(\sigma_1')\}
=4\pi\delta^\oi_\oj\delta(\sigma_1-\sigma_1')\,,\qquad
\{\bpsi^i(\sigma_1),\bpsi_j(\sigma_1')\}
=4\pi\delta^i_j\delta(\sigma_1-\sigma_1')\,.
$$
The supersymmetry generators are
\be\label{eqn:QA}
Q=\int \psi^\oi\bigl(
\Pi_\oi + G_{i\oi}\partial_1\smX^i
\bigr)d\sigma_1\,,\qquad
\bQ=\int \bpsi_i\bigl(
G^{i\oi}\Pi_\oi + \partial_1\smX^i + \cdots
\bigr)d\sigma_1\,
\ee
[where $(\cdots)$ denotes higher order terms that are not important here],
so that 
$$
\delta(\cdots)=i[\pSUSY Q+\bpSUSY\bQ,(\cdots)].
$$
Note that the definition of $Q$ does not suffer from any normal ordering ambiguity, while the definition of $\bQ$ does require a normal ordering prescription (for $G_{i\oj}(\smX)$ and $\Pi_\oi$). We take this opportunity to recall that the sigma-model can be topologically twisted \cite{Witten:1991zz}, and for our flat worldsheet ($S^1\times\R$) the twisting has no effect on the spectrum. With ordinary supersymmetric (Ramond-Ramond) boundary conditions on $S^1$ the ground states correspond to Dolbeault cohomology, and $Q$ is identified with $\bpartial$ \cite{Witten:1991zz}. In the rest of this paper we will look for a geometric interpretation for the ground states of the model with boundary conditions that include a mirror-symmetry twist.

\section{$T^2$ target space}
\label{sec:T2}

The simplest target space we can consider is $\TGT=T^2.$ 
We parameterize the metric and $B$-field in terms of two complex coordinates $\rho\equiv\rho_1+i\rho_2$ and $\tau\equiv\tau_1+i\tau_2$ as
$$
ds^2\equiv G_{IJ}d\ttX^I d\ttX^J
= \frac{\rho_2}{\tau_2}|d\ttX^1+\tau d\ttX^2|^2\,,\qquad
B_{IJ}d\ttX^I\wedge d\ttX^J = 2\rho_1 d\ttX^1\wedge d\ttX^2\,,
$$
where $0\le \ttX^1,\ttX^2<2\pi$ are periodic coordinates.
The action is (and we now switch from coordinates $\ttX^I$ to fields $\phi^I$):
\bear
I_0 &=&\tfrac{1}{2\pi}\int\bigl(
(G_{IJ}+i B_{IJ})\partial\smX^I\bpartial\smX^J
+i G_{IJ}\psi^I\partial\psi^J
+i G_{IJ}\bpsi^I\bpartial\,\bpsi^J
\bigr)d^2\sz\,.
\label{eqn:SMT2}
\eear
T-duality acts as $\rho\rightarrow -1/\rho$, keeping $\tau$ fixed, and is a symmetry for $\rho=i.$ From now on we assume that we are at the self-dual point $\rho=i$ of moduli space. Now, let us compactify direction $\sigma^1$ on $S^1$ so that $0\le\sigma^1<2\pi.$
Usual (Ramond-Ramond) periodic boundary conditions require that the value of every fermionic field at $\sigma^1=0$ be equal to its value at $\sigma^1=2\pi$, and for bosonic fields both field and first derivative should be equal.
In order to introduce a T-duality twist we first relax the condition of continuity and allow the fields at $\sigma^1=0$ to be independent of the fields at $\sigma^1=2\pi$. We denote by a subscript $\sgt$ or $\slt$ the limiting values of the various fields as $\sigma^1\rightarrow 0$ or $\sigma^1\rightarrow 2\pi$:
\be\label{eqn:Xpm}
\left.\begin{array}{ll}
\smX^I_\sgt\equiv\smX^I(0,\sigma^2)\,,&
\smX^I_\slt\equiv\smX^I(2\pi,\sigma^2)\,,\\
\psi^I_\sgt\equiv\psi^I(0,\sigma^2)\,, &
\psi^I_\slt\equiv\psi^I(2\pi,\sigma^2)\,,\\
\bpsi^I_\sgt\equiv\bpsi^I(0,\sigma^2)\,,&
\bpsi^I_\slt\equiv\bpsi^I(2\pi,\sigma^2)\,. \\
\end{array}\right\}
\ee
The T-duality twist is then accomplished by adding the following term to the action $I_0$ of \eqref{eqn:SMT2}:
\be\label{eqn:I1T}
I_1=\frac{i}{2\pi}\int\sqrt{G}
\epsilon_{IJ}\left\lbrack\smX^I_{\slt}\partial_2\smX^J_{\sgt}
+2\psi^I_{\slt}\psi^J_{\sgt}+2\bpsi^I_{\slt}\bpsi^J_{\sgt}
\right\rbrack d\sigma^2
\,,
\ee
where $\partial_2\equiv\partial/\partial\sigma^2$.
The classical equations of motion that one gets by varying the action $I_0+I_1$, where $I_0$ is given in \eqref{eqn:SMT2} and $I_1$ in \eqref{eqn:I1T}, are the free field equations away from the twist point $\sigma^1=0$, and at the twist point we get the boundary conditions:
\be\label{eqn:Tbc}
{\epsilon^I}_J\partial_2\smX^J_{\sgt} = i\partial_1\smX^I_{\slt}
\,,\qquad
{\epsilon^I}_J\partial_1\smX^J_{\sgt} = -i\partial_2\smX^I_{\slt}
\,,\qquad
{\epsilon^I}_J\psi^J_{\sgt} = -\psi^I_{\slt}
\,,\qquad
{\epsilon^I}_J\bpsi^J_{\sgt} = -\bpsi^I_{\slt}
\,.
\ee
These equations indeed realize T-duality.
The term $I_1$ is also invariant under \SUSY{(2,2)} supersymmetry with parameters $\pSUSY,\bpSUSY$ (using the bulk Dirac equation):
\bear
\delta\smX^I &=& i\bpSUSY\psi^I+i\pSUSY\bpsi^I+i\bpSUSY'{\epsilon^I}_J\psi^J+i\pSUSY'{\epsilon^I}_J\bpsi^J\,,
\nn\\
\delta\psi^I &=& -\bpSUSY\partial_\sz\smX^I-\bpSUSY'{\epsilon^I}_J\partial_\sz\smX^J\,,
\nn\\
\delta\bpsi^I &=& -\pSUSY\partial_\bsz\smX^I-\pSUSY'{\epsilon^I}_J\partial_\bsz\smX^J\,.
\nn
\eear
Beyond the classical level, the identification of \eqref{eqn:I1T} with the T-duality twist can be argued by taking direction $\sigma^1$ to be (Euclidean) time and realizing the operator of T-duality on wave-functionals of the fields at $\sigma^1=0$. The result is that a wave-functional of the field $\smX^I_{\slt}(\sigma^1)$ can be converted to the dual wave-functional by multiplying the wave-functional by the bosonic part of $e^{-I_1}$ and path-integrating over $[\cD\smX^I_{\slt}]$. (See \cite{Ganor:2008hd} for more details.)

\subsection{Discrete symmetries}
\label{subsec:DiscSymmT2}
Target space momentum and winding number are not conserved by $I_1$ [defined in \eqref{eqn:I1T}] but there is a combination of them that is conserved and gives rise to a useful discrete symmetry. To find this combination, consider first the theory without the T-duality twist. The momentum $(p_1, p_2)$ and winding $(w^1,w^2)$ quantum numbers (which take integer values) are conserved, but they do not commute with T-duality, which acts on them as:
$$
\begin{pmatrix} p_1 \\ p_2 \\ w^1 \\ w^2 \\ \end{pmatrix}\rightarrow
\begin{pmatrix} w^2 \\ -w^1 \\ p_2 \\ -p_1 \\ \end{pmatrix}.
$$
However, the combinations
$$
\symPW_1\equiv (-1)^{p_1 + w^2}\,,\qquad
\symPW_2\equiv (-1)^{p_2 + w^1}\,,
$$
which take values in $\Z_2$, are conserved.
We will argue below in \secref{subsec:GroundStatesT2} that when we insert the T-duality twist, $\symPW_1$ and $\symPW_2$ give rise to good $\Z_2$-valued quantum numbers, but they acquire a nontrivial commutation relation:
\be\label{eqn:symPWcom}
\symPW_1\symPW_2 = -\symPW_2\symPW_1\,.
\ee

\subsection{Mode expansion}
\label{subsec:ModeExpansionT2}
T-duality acts as 
$$
\partial\phi^I\rightarrow G^{IJ}\epsilon_{JK}\partial\phi^K
\,,\qquad
\bpartial\phi^I\rightarrow -G^{IJ}\epsilon_{JK}\bpartial\phi^K
\,,
$$
or explicitly,
$$
\partial\phi^\One + \tau\partial\phi^\Two\rightarrow
i(\partial\phi^\One + \tau\partial\phi^\Two)
\,,\quad
\partial\phi^\One + \btau\partial\phi^\Two\rightarrow
-i(\partial\phi^\One + \btau\partial\phi^\Two)
$$
and similarly
$$
\bpartial\phi^\One + \tau\bpartial\phi^\Two\rightarrow
-i(\bpartial\phi^\One + \tau\bpartial\phi^\Two)
\,,\quad
\bpartial\phi^\One + \btau\bpartial\phi^\Two\rightarrow
i(\bpartial\phi^\One + \btau\bpartial\phi^\Two)
\,.
$$
With the T-duality twist, the mode expansion is:
\bear
\psi^\One +(\tau_1\pm i\tau_2)\psi^\Two &=&
\sum\limits_{n\in\Z}
\psiMode^\pm_{n\pm \frac{1}{4}}e^{i (n\pm \frac{1}{4})(\sigma_1+i\sigma_2)}
\,,\quad
\bpsi^\One +(\tau_1\pm i\tau_2)\bpsi^\Two = 
\sum\limits_{n\in\Z}
\bpsiMode^\pm_{n\mp \frac{1}{4}}e^{i (n\mp \frac{1}{4})(-\sigma_1+i\sigma_2)}\,,
\nn\\ &&
\label{eqn:PsiModes}
\eear
and,
\bear
\partial\smX^\One +(\tau_1\pm i\tau_2)\partial\smX^\Two &=&
\sum\limits_{n\in\Z}
\phiMode^\pm_{n\pm\tfrac{1}{4}} e^{i (n\pm\tfrac{1}{4})(\sigma_1+i\sigma_2)}
\,,\quad
\bpartial\smX^\One +(\tau_1\pm i\tau_2)\bpartial\smX^\Two = 
\sum\limits_{n\in\Z}
\wphiMode^\pm_{n\mp\tfrac{1}{4}} e^{i (n\mp\tfrac{1}{4})(-\sigma_1 + i\sigma_2)}
\,.\nn\\ &&
\label{eqn:PhiModes}
\eear
Thus, the boundary conditions \eqref{eqn:Tbc} create a spectral flow in the free field mode expansion and the modes are shifted to values in $\Z\pm\frac{1}{4}$.

As we shall see in \secref{subsec:GroundStatesT2}, there are two ground states. We denote them by $\ket{j}$, with $j=0,1$. They satisfy
$$
0=
\phiMode^{+}_{n-\frac{1}{4}}\ket{j}=\wphiMode^{-}_{n-\frac{3}{4}}\ket{j}=
\phiMode^{-}_{n-\frac{3}{4}}\ket{j}=\wphiMode^{+}_{n-\frac{1}{4}}\ket{j}\,,
\qquad
n=1,2,3,\dots
$$
and
$$
0=
\psiMode^{+}_{n-\frac{1}{4}}\ket{j}=\bpsiMode^{-}_{n-\frac{3}{4}}\ket{j}=
\psiMode^{-}_{n-\frac{3}{4}}\ket{j}=\bpsiMode^{+}_{n-\frac{1}{4}}\ket{j}\,,
\qquad
n=1,2,3,\dots
$$

\subsection{Ground states}
\label{subsec:GroundStatesT2}

The mode expansions \eqref{eqn:PsiModes}-\eqref{eqn:PhiModes} show that after the T-duality twist the worldsheet fields no longer have any zero modes. Nevertheless, we will now argue that the ground state is a doublet.
One way to see this is by performing T-duality on direction $\ttX^1$.
This duality interchanges the moduli $\rho$ and $\tau$ and replaces the T-duality twist (which corresponded to $\rho\rightarrow -1/\rho$) with an ordinary geometrical twist ($\tau\rightarrow -1/\tau$) which for $\tau=i$ corresponds to a rotation by $\pi/2$ of the $T^2$ (around some fixed point, say the origin). We denote the scalar fields of this dual worldsheet theory by $\tPhi^1$ and $\Phi^2$, so that $\tPhi^1+i\Phi^2$ is a coordinate on the dual torus.
In this T-dual description the ground states correspond to fixed points of the rotation. These are the two allowed discrete values of the zero mode of the (dual) scalar worldsheet field, and there are two such fixed points: either the origin ($\tPhi^1=\Phi^2=0$) or the center of the torus ($\tPhi^1=\Phi^2=\pi$). Furthermore, the operators $\symPW_1$ and $\symPW_2$ acquire a geometrical interpretation.
$\symPW_1$ is mapped to a translation (which commutes with the twist):
$$
\symPW_1:\qquad\tPhi^1+i\Phi^2\mapsto(\tPhi^1+\pi)+i(\Phi^2+\pi)\,.
$$
The operator $\symPW_2$ is mapped to a topological charge which can be defined as follows (see \cite{Ganor:2010md} for a related discussion). Consider the $3$-dimensional space $M$ that we construct as a $T^2$ fibration over $S^1$ with the fiber being the target space and the base being the $\sigma^1$ circle. The homology of this space is $H_1(M,\Z)=\Z\oplus\Z_2$ where the $\Z$ factor is generated by the homology class of a loop that wraps around the $S^1$ base at constant $T^2$ position $\tPhi^1=\Phi^2=0$, and the $\Z_2$ factor is generated by the homology class of a loop that wraps one of the short nontrivial cycles of the fiber $T^2$ at a constant $\sigma^1$ (which becomes a torsion class because of the identification induced by the twist). Now let $(\tPhi^1(\sigma^1),\Phi^2(\sigma^1))$ be any smooth field configuration that respects the twist. We can associate a topological $\Z_2$ charge with it by interpreting it as a loop in $M$. The $\Z_2$ charge (with values $\pm 1$) is then determined from the homology of the loop. This $\pm 1$ charge is then identified with the eigenvalue of the operator $\symPW_2$. It is also not hard to check (see \cite{Ganor:2010md}) that the loop at the origin ($\tPhi^1=\Phi^2=0$) and the loop at the center ($\tPhi^1=\Phi^2=\pi$) have opposite $\symPW_2$ charge. If we denote by $\ket{0}$ the ground state that corresponds to the mode expansion around $\tPhi^1=\Phi^2=0$, and by $\ket{1}$ the ground state that corresponds to the mode expansion around $\tPhi^1=\Phi^2=\pi$ then we can summarize the above observations as
$$
\symPW_1\ket{0}=\ket{1}\,,\qquad
\symPW_1\ket{1}=\ket{0}\,,\qquad
\symPW_2\ket{0}=\ket{0}\,,\qquad
\symPW_2\ket{1}=-\ket{1}\,,
$$
where we have arbitrarily assigned to $\ket{0}$ the $\Z_2$ charge $+1$.
In particular, the commutation relation \eqref{eqn:symPWcom} follows.
In section \secref{subsec:relgq} we will present another way to verify that there are two ground states, in terms of deformation quantization of the target space.

\subsection{Fermion number}
\label{subsec:F}
The fermions have modes labeled by $n\pm\frac{1}{4}$, and all the zero modes have been eliminated by the T-duality twist. However, the ground states $\ket{j}$ have odd fermion number $(-1)^F= -1$, as we will now argue.
To see this, we compactify the worldsheet on $T^2$ and calculate the Witten index in two ways. From the discussion above, the Witten index is $2(-1)^F$.
On the other hand, by modular invariance, the Witten index is also 
$$
I = \tr_{\Hilb_0}\left\lbrack (-1)^F\mvS\right\rbrack,
$$
where the trace is taken over the Hilbert space of ground states of the 1+1D $\sigma$-model compactified on $S^1$ {\it without} a T-duality twist, and $\mvS$ is the T-duality operator on that Hilbert space.
The T-duality operator is part of the $\SL(2,\Z)_\rho\times\SL(2,\Z)_\tau$ group of dualities that act on the Hilbert space of ground states. Here $\SL(2,\Z)_\rho$ acts on the complexified area of $T^2$, while $\SL(2,\Z)_\tau$ is the mapping class group which acts on the complex structure $\tau$. The ground states of $\Hilb_0$ correspond to the cohomology of $T^2$ and are in the $(2,1)\oplus (1,2)$ representation of the duality group. The two states corresponding to $H^1(T^2,\Z)$ have odd fermion number and are singlets of $\SL(2,\Z)_\rho$ while the two states of $H^0(T^2,\Z)\oplus H^2(T^2,\Z)$ are a doublet of $\SL(2,\Z)_\rho$ and a singlet of $\SL(2,\Z)_\tau$. The operator $\mvS$ acts as the identity on $H^1(T^2,\Z)$ and as the matrix $\begin{pmatrix} 0 & 1 \\ -1 & 0 \\ \end{pmatrix}$ on $H^0(T^2,\Z)\oplus H^2(T^2,\Z)$. The Witten index is therefore $-2$ and the states $\ket{j}$ therefore have odd fermion number, as stated.

\subsection{Other elements of $\SL(2,\Z)$}
\label{subsec:OtherSL}

We can also consider twisting the boundary conditions by the $\SL(2,\Z)$ transformations
$$
\pm
\begin{pmatrix}
1 & -1 \\ 1 & 0 \\
\end{pmatrix}
$$
which keep the $\rho=e^{\pi i/3}$ value fixed and generate an abelian group of order $\ord=6$ (for the $+$ sign) and $\ord=3$ (for the $-$ sign).
In addition, it is also useful to consider as a test case the twist by the central element $-1\in\SL(2,\Z)$. This is equivalent to a geometrical rotation by $\pi$ and keeps every value of $\rho$ fixed. It has order $\ord=2$.
The modified mode expansions have modes with values in $\Z\pm\tfrac{1}{\ord}$.
There are $\lvk$ ground states $\ket{j}$ ($j=0,\dots,\lvk-1$) with $\lvk=1$ for $\ord=6$, $\lvk=3$ for $\ord=3$ and $\lvk=4$ for $\ord=2$. Further explanation of the meaning of $\lvk$ follows in section \secref{subsec:relgq} and \secref{sec:ConnectGQ}.

\subsection{Relation to the noncommutative torus and the Landau problem}
\label{subsec:relgq}
We are interested in the ground states of the theory described by the action $I_0+I_1$ [see \eqref{eqn:SMT2} and \eqref{eqn:I1T}].
The ground states of this system can also be established by a standard analysis of the relevant terms at low-energy. We have compactified the 1+1D theory on $S^1$, and below the Kaluza-Klein scale (the inverse of the radius $\xL_1$ of the $\sigma^1$ circle) we can neglect the dependence of the fields on $\sigma_1$.
More precisely, the boundary conditions \eqref{eqn:Tbc} imply that excited states have energy of the order of the $\sigma^1$-compactification scale (at least $1/4\xL_1$). The only terms that survive the low-energy limit are those coming from $I_1$, and since there is no longer any $\sigma^1$ dependence, we can set $\smX^I_{\sgt}=\smX^I_{\slt}\equiv\smX^I(\sigma^2)$
, and similarly for the fermionic fields. The fermionic fields become massive and irrelevant at low-energy, and we are left with the 0+1D action
\be\label{eqn:I1LE}
I=-\frac{i}{2\pi}\int
\epsilon_{IJ}\smX^I\partial_2\smX^J
d\sigma^2
=-\frac{i}{2\pi}\int \epsilon_{IJ}\smX^I d\smX^J
\,.
\ee
Note that we have dropped the kinetic terms proportional to $(\partial_2\smX^I)^2$ since they are IR irrelevant, as can be verified by simple dimensional analysis. It is nevertheless convenient to add the kinetic term again so as to get the Lagrangian of a Landau problem (a particle in a uniform magnetic field) on $T^2$:
\be\label{eqn:LLag}
I_L=\frac{1}{2\pi}
\int\left(\tfrac{1}{2}m G_{IJ}\dot{\xf}^I\dot{\xf}^J - \frac{i\lvk}{2}\epsilon_{IJ}\xf^I\dot{\xf}^J\right)dt\,,
\ee
where we set $t\equiv\sigma^2$, and $m=2\pi\xL_1$ is the mass parameter (which is the circumference of the $\sigma^1$ direction). We have also introduced $\lvk\equiv 2$ in \eqref{eqn:LLag}. Other values of $\lvk$ appear in connection with the other elements of the $\SL(2,\Z)$ duality group discussed in \secref{subsec:OtherSL}.
The ground states of \eqref{eqn:LLag} are Landau wave-functions (independent of $m$ and the area of the $T^2$):
\bear
 \wfF_{j,\lvk}(\xf_1,\xf_2) &=&
\frac{1}{2\pi}
\left(2 \lvk \tau_2 \right)^{\frac{1}{4} } 
e^{\frac{i\lvk\xf_1\xf_2}{4\pi}}
\sum_{n=-\infty}^\infty e^{i(\lvk n+j)\xf_1+ \pi i\lvk\tau (\frac{\xf_2}{2\pi}+n+\frac{j}{\lvk})^2}
\nn\\ &=&
\frac{1}{2\pi}
\left(2 \lvk \tau_2 \right)^{\frac{1}{4} } 
e^{\frac{i\lvk\xf_1\xf_2}{4\pi}}
 e^{ \pi i \tau \lvk \left( \tfrac{ \xf_2 }{2 \pi}  \right)^2}  \Theta _{j,\lvk} (\tfrac{\xf_1 +\tau\xf_2}{2\pi}; \tau )
\label{eqn:Lwf}
\eear
where the $\Theta$-function is defined as
\be\label{eqn:Thetafn}
\Theta_{j,\lvk}(u,\tau)\equiv
\sum_{n=-\infty}^\infty e^{\pi i\lvk\tau(n+\frac{j}{\lvk})^2 
+ 2\pi i\lvk(n + \frac{j}{\lvk})u}
\ee
and is holomorphic in $u$ and $\tau$.

In the low-energy limit the kinetic term in \eqref{eqn:LLag} can be dropped, and as is well-known, we are left with the Lagrangian that describes a noncommutative $T^2$ with symplectic form $\frac{\lvk}{4\pi^2}d\xf^1\wedge d\xf^2$. Thus, $\lvk$ has the interpretation of the level of the deformation quantization problem of $T^2$. We will return to this point in \secref{sec:ConnectGQ}.

\section{Elliptically fibered $K_3$ target spaces}
\label{sec:EFK3}

The analysis of \secref{sec:T2} can be extended to curved target spaces $\TGT$ that can be represented as an {\it elliptic fibration}. Such spaces are described by fibering a torus with a complex structure that varies over some base $\efBase$. More precisely, there exists a surjective holomorphic map $\pi$ from the $\dimTGT$ dimensional space $\TGT$ to a complex space $\efBase$ of (complex) dimension $\dimTGT-1$ such that the inverse image $\pi^{-1}(p)$ of a generic point $p\in\efBase$ is a torus. 
The complex structure $\tau$ of the fiber $\pi^{-1}(p)$ is allowed to vary (holomorphically) with the point $p$ along the base $\efBase.$ An important example is $\TGT=K_3$ with $\efBase=\CP^1.$ This construction was developed in detail in \cite{Greene:1989ya} and  is central to F-theory \cite{Vafa:1996xn}. 
The complex structure $\tau$ is allowed to become $\infty$ (possibly after an $\SL(2,\Z)$ transformation) at a codimension-$1$ submanifold of $\efBase$, which for a generic elliptically fibered $K_3$ turns out to be $24$ isolated points $\zB_1,\zB_2,\dots,\zB_{24}$. We will assume that sufficiently far from these points $\tau$ varies very slowly over distances of order $1$ (the string-scale). The metric then takes the approximate form
\be\label{eqn:MetricEF}
ds^2 = \frac{\rho_2}{\tau_2}|d\xf_1 + \tau d\xf_2|^2
+g_{\zB\bzB}|d\zB|^2
\,,
\ee
where $\zB$ is a complex coordinate on $\efBase$, $0\le \xf_1, \xf_2<2\pi$ are periodic coordinates on the fiber, $\rho_2$ is the constant area of the fiber, $\tau(\zB)$ is a locally defined holomorphic function of the base, and $\tau_2$ is the imaginary part of $\tau.$
The behavior of the complex structure $\tau$ as a function of $\zB$ is conveniently encoded by the equation \cite{Greene:1989ya}:
\be\label{eqn:efxy}
y^2 = x^3 - f(\zB)x-g(\zB)\,,
\ee
where $f$ and $g$ are holomorphic functions of $\zB$, which for a $K_3$ are polynomials of degrees $8$ and $12$, respectively.
The complex structure is given in terms of the $j$-function
$$
j(\tau)=\frac{64f^3}{\frac{f^3}{27}-\frac{g^2}{4}}\,.
$$
The hyper-K\"ahler metric $g_{\zB\bzB}$ is determined as follows.
In the notation of \cite{Seiberg:1994rs}, one defines a holomorphic function $a(\zB)$ by solving $\frac{da}{d\zB} = \oint_\alpha\frac{dx}{y}$, where $\alpha$ is a basic $1$-cycle of the $T^2$ fiber at $\zB$ (the cycle that corresponds to a constant $\xf_2$, with $\xf_1$ running from $0$ to $2\pi$). We then have
$g_{\zB\bzB}=\tau_2\left|\frac{da}{d\zB}\right|^2.$
The form of the metric \eqref{eqn:MetricEF}, where the components are independent of $\xf_1$ and $\xf_2$, is approximate, with corrections that are exponentially small sufficiently far away from the points $\zB_i$ of the singular fibers.\footnote{
In Seiberg-Witten theory such spaces appear as moduli spaces of 3+1D $SU(2)$ gauge theories compactified on $S^1$ \cite{Seiberg:1996nz}. We are here considering the limit that the Kaluza-Klein energy scale is much smaller than the QCD scale, and the exponentially small corrections arise from BPS particles with Euclidean worldline along the circle.}

Now we take a sigma-model with a metric of the approximate form \eqref{eqn:MetricEF}. We set $\rho=i$ ($\rho$ is part of the K\"ahler moduli of the sigma-model). T-duality of the fibers now extends to mirror-symmetry of the whole space \cite{Strominger:1996it,Hori:2000kt}. We can then compactify on $S^1$ with a mirror-symmetry twist.
The goal of this section is to understand the ground states of the twisted compactification in terms of a semi-classical description involving $\efBase$ alone.
We specialize to the $K_3$ case for which, as mentioned above, there are $24$ points $\zB_1,\dots,\zB_{24}$ where the fiber degenerates. We will refer to them as {\it special points.} We have chosen a metric on the base that is large enough so that (sufficiently far away from the special points) the complex structure $\tau(\zB)$ varies sufficiently slowly, and we can then study the problem in a Born-Oppenheimer approximation whereby we treat the fields along the fiber as the ``fast modes'' and the fields along the base as the ``slow-modes.''
As we will describe below, we can then reduce the problem to what is essentially a quantum mechanical problem with $\efBase$ as the configuration space.
We will develop this quantum mechanical description along the following lines:
\begin{enumerate}
\item
Sufficiently far away from the special points $\zB_i$, we can derive a simple wave-equation for the wave-function.

\item
The wave-functions have nontrivial monodromies around special points. In general when passing through a cut that emanates from a special point, the $T^2$ fiber undergoes a certain $\SL(2,\Z)$ (mapping class group) transformation, which acts nontrivially on the wave-function \cite{Greene:1989ya, Vafa:1996xn}.
These nontrivial monodromies around the $\zB_i$'s can be eliminated by encoding the wave-functions in terms of sections of certain holomorphic line bundles on the total space. The boundary conditions at the special points are determined by normalizability of the wave-function.

\item
There are additional ``bound states'' whose wave-functions are localized at the special points and are out of reach of the classical analysis above.
However, the Witten index allows us to glean some information about these bound states.

\end{enumerate}
We now proceed to describe each part in detail.

\subsection{Supersymmetry}
\label{subsec:SUSY}
The moduli space of the $\sigma$-model with $K_3$ target space is \cite{Seiberg:1988pf,Aspinwall:1994rg}:
\be\label{eqn:ModSpaceO}
O(20,4,\Z)\backslash O(20,4,\R)/(O(20)\times O(4))
=O^+(20,4,\Z)\backslash O^+(20,4,\R)/(O(20)\times SO(4)),
\ee
where $O^+(20,4,\R)$ denotes an index-$2$ subgroup of $O(20,4,\R).$
The amount of worldsheet supersymmetry in the mirror-twisted setting is determined as follows. The $\sigma$-model has \SUSY{(4,4)} supersymmetry with an $SU(2)_L\otimes SU(2)_R$ R-symmetry (with $SU(2)_L$ acting on the left-moving sector of the CFT and $SU(2)_R$ acting on the right-moving sector). The supercharges are in the spinor representation $(2,1)\oplus(1,2)$.
The R-symmetry $SU(2)_L\otimes SU(2)_R$ can be identified with a double-cover of the $SO(4)$ factor in \eqref{eqn:ModSpaceO} in the following sense \cite{Seiberg:1988pf,Aspinwall:1994rg}. 
Set 
$$
K\equiv O(20)\times SO(4)\,,\qquad
G\equiv O^+(20,4,\R)\,,\qquad
\Gamma\equiv O^+(20,4,\Z)\,.
$$
Then, the supercharges of the theory live in a vector bundle over the moduli space $\Gamma\backslash G/K$ whose structure group is $SU(2)_L\times SU(2)_R$ which is a double cover of $SO(4)$. The principal bundle [i.e., the bundle one gets upon replacing the fiber of the vector bundle with the structure group $SO(4)$] is $\Gamma\backslash G/SO(20)$ (which is an $SO(4)$ bundle over the base $\Gamma\backslash G/K$). 
Now let $gK\in G/K$ be the point in moduli space, represented as a coset of groups. If $\OpM\in O^+(20,4,\Z)$ preserves $gK$ then $g^{-1}\OpM g\in K.$ Let $\rho\in SO(4)$ be the projection of $g^{-1}\OpM g$ to the $SO(4)$ factor of $K$. We will now argue that the left-moving supercharges transform as $(1,1)\oplus (3,1)$ under $\rho\in SO(4)$ and the right-moving supercharges transform as $(1,1)\oplus(1,3)$ [where representations of $SO(4)$ are written as representations of $SU(2)_L\otimes SU(2)_R$]. The singlets correspond to the preserved supersymmetries \eqref{eqn:SusyI}, while the triplets can be understood as follows. Schematically, the left-moving supercurrents are of the form $\psi\partial\smX$. The R-symmetries act on the $\psi$'s but not the $\smX$'s, but $\rho$ needs to act on $\partial\smX$ in a dual way to $\psi$ in order to commute with the supersymmetries of $\eqref{eqn:SusyI}$.
Assuming that $\psi$ transforms in $(2,1)$, we find that the SUSY generators are in $(2,1)\otimes (2,1)=(1,1)\oplus (3,1)$ as stated above.

As a special case, consider a point in moduli space that is invariant under an isometry. The isometry corresponds to an element $\OpM'\in O^+(19,3,\Z)\subset O^+(20,4,\Z)$ and the projection $\rho'$ of $g^{-1}\OpM' g$ to the $SO(3)\subset SO(4)$ factor of $K$ determines the amount of preserved supersymmetry as follows. The isometry always preserves the supersymmetries of \eqref{eqn:SusyI}, and for every complex structure that the isometry preserves we get an extra supersymmetry of the type \eqref{eqn:SusyII}. By construction \cite{Aspinwall:1994rg}, the complex structures are in a triplet of $SO(3)$ (the group of which $\rho'$ is an element), and so the number of (left and right) preserved supersymmetries of type \eqref{eqn:SusyII} is the number of eigenvalues of $1$ of $\rho'$. As an element of $SO(4)$, $\rho'$ has an extra eigenvalue of $1$, which corresponds to the preserved supersymmetry of type \eqref{eqn:SusyI}.

\subsection{Perturbative analysis}
\label{subsec:PertAnal}

We will now construct the wave equation that the ground-state wave-functions satisfy for a target space constructed by fibering a $T^2$ (of complex structure $\tau$) over an open neighborhood of $\C$ (parameterized by $\zB$), with a slowly varying $\tau(\zB)$.
We take the metric as in \eqref{eqn:MetricEF} and choose the fields that correspond to the complex coordinates as follows:
\be\label{eqn:chgxfzF}
\zB\rightarrow\smX^\zi\,,\quad
\bzB\rightarrow\smX^\bzi\,,\quad
\zF\equiv\tfrac{1}{2\pi}(\xf_1+\tau(\zB)\xf_2)\rightarrow\smX^\wi\,,\quad
\bzF\equiv\tfrac{1}{2\pi}(\xf_1+\btau(\bzB)\xf_2)\rightarrow\smX^\bwi\,.
\ee
The periodicity of the holomorphic coordinate $\zF$ is $\zF\sim\zF+1\sim\zF+\tau$, but the dependence of $\tau$ on $\zB$ is inconvenient here. We will therefore work in a mixed convention where the fundamental fields are
$$
\smX^\zi\,,\quad
\smX^\bzi\,,\quad
\smX^1\,,\quad
\smX^2\,,\quad
\psi^\bzi\,,\quad
\psi_\bzi\,,\quad
\bpsi_\zi\,,\quad
\bpsi^\zi\,,\quad
\psi^1\,,\quad
\psi^2\,,\quad
\bpsi^1\,,\quad
\bpsi^2\,.
$$
We denote the canonical dual momenta of the bosons in this formalism by
$$
\canMom_\zi'\,,\quad
\canMom_\bzi'\,,\quad
\canMom_1'\,,\quad
\canMom_2'\,.
$$
They are related to the canonical momenta of $\smX^\wi$ and $\smX^\zi$ by\footnote{The following relations are obtained by applying the change of variables from $(\xf_1, \xf_2, \zB,\bzB)$ to $(\zF, \bzF, \zB,\bzB)$ of \eqref{eqn:chgxfzF}. For example, $\canMom_\wi$ transforms as a component of a $1$-form, and $\psi^\bwi$ as a component of a vector.}
$$
\canMom_\wi=\frac{i}{2\tau_2}(\btau\canMom_1'-\canMom_2')
\,,\qquad
\canMom_\zi=\canMom_\zi'
+\frac{i\tau'\smX^2}{2\tau_2}(\canMom_2'-\btau\canMom_1')\,,
$$
and $\psi^1$ and $\psi^2$ are related to $\psi^\bwi$ and $\psi^\bzi$ by:
$$
\psi^\bwi = \psi^1+\btau\psi^2+\btau'\smX^2\psi^\bzi\,.
$$
We rewrite the left equation in \eqref{eqn:QA} as
\be\label{eqn:QAii}
Q=\int \left\lbrack
\psi^\bzi\bigl(
\Pi_\bzi' + G_{i\bzi}\partial_1\smX^i
\bigr)
+\frac{i}{2\tau_2}(\psi^1+\btau\psi^2)\bigl(
\canMom_2'-\tau\canMom_1'
+G_{2 I}\partial_1\smX^I
-\tau G_{1 I}\partial_1\smX^I
\bigr)
\right\rbrack
d\sigma_1\,,
\ee
Now, we put the theory on $S^1$ with a mirror-symmetry twist.
This is achieved by adding a term similar to \eqref{eqn:I1T} to the action.
The mirror-symmetry operation acts as
$$
(\psi^1+\btau\psi^2)\rightarrow i(\psi^1+\btau\psi^2)\,,
$$
and in order to preserve $Q$ it must also act as,
$$
\bigl(
\canMom_2'-\tau\canMom_1'
+G_{2 I}\partial_1\smX^I
-\tau G_{1 I}\partial_1\smX^I
\bigr)\rightarrow 
-i \bigl(
\canMom_2'-\tau\canMom_1'
+G_{2 I}\partial_1\smX^I
-\tau G_{1 I}\partial_1\smX^I
\bigr)\,.
$$
The mirror-symmetry twist eliminates the zero-modes of $(\psi^1+\btau\psi^2)$ and of the term it multiplies in \eqref{eqn:QAii}, and their modes become nonintegral (in $\Z\pm\frac{1}{4}$). Since we are interested in the ground states, in the limit of slowly varying $\tau(\zB)$ we can ignore the terms that contain these fields in \eqref{eqn:QAii}. We are therefore left with
\be\label{eqn:Qapprox}
Q\rightarrow\int
\psi^\bzi\bigl(
\Pi_\bzi' + G_{i\bzi}\partial_1\smX^i
\bigr)
d\sigma_1
\rightarrow
\int\psi^\bzi\Pi_\bzi'
d\sigma_1\,.
\ee
We also dropped the term proportional to $\partial_1\smX^i$ because we are only interested in modes with zero momentum along $S^1$.
The ground states can now be constructed as follows.
Let $\bpsiMode^\zi_0$ and $\psiMode^\bzi_0$ be the zero modes in the mode expansion of $\bpsi^\zi$ and $\psi^\bzi$, respectively.
For constant $\tau(\zB)$ (i.e., for $\C\times T^2$) we build a basis of ground state wave-functions by combining a ground state of the $T^2$ compactification with a ground state of the $\C$ compactification. The ground states of the $T^2$ compactification are labeled by $j=0,\dots,\lvk-1=1$, as in \secref{subsec:GroundStatesT2} (and we keep track of $\lvk$ for possible generalizations as in \secref{subsec:OtherSL}). The ground states of the $\C$ problem correspond to Dolbeault $\bpartial$-cohomology on $\C$ (which are not normalizable, but that is not going to be a problem since we are not going to use all of $\C$ but rather we are going to patch open subsets of $\C$ to form a $\CP^1$). We denote by $\ket{j}$ the state that satisfies
$\bpsiMode_{\zi\,0}\ket{j}=\psiMode_{\zi\,0}\ket{j}=0$ and corresponds to the $j^{th}$ ground state of the $T^2$ problem.
A generic ground state can then be written as
$$
\omega_j(\zB,\bzB)\ket{j}
+\omega_{j,\zi}(\zB,\bzB)\bpsiMode^\zi_0\ket{j}
+\omega_{j,\bzi}(\zB,\bzB)\psiMode^\bzi_0\ket{j}
+\omega_{j,\zi\bzi}(\zB,\bzB)\bpsiMode^\zi_0\psiMode^\bzi_0\ket{j}
$$
where $(\zB,\bzB)$ are the ``center-of-mass'' coordinates in the $\zB$ (and $\bzB$) directions.
To be annihilated by $Q$ requires
$$
\chi_j\equiv
\omega_j+
\omega_{j,\zi} d\zB+
\omega_{j,\bzi} d\bzB+
\omega_{j,\zi\bzi}d\zB\wedge d\bzB
$$
to be a sum of $d\bzi\bpartial_\bzi$-closed forms, and in the topologically twisted theory (the A-model as in \cite{Witten:1991zz}) we also impose an equivalence up to $d\bzi\bpartial_\bzi$-exact forms. Since the theory is on a flat worldsheet, the topologically twisted theory is equivalent to the untwisted one, and so $\chi_j$ is in the $d\bzi\bpartial_\bzi$-cohomology. In particular, $\omega_j$ is a holomorphic $0$-form and $\omega_{j,\zi}d\zB$ is a holomorphic $1$-form.
Given the structure of $Q$ and the approximation \eqref{eqn:Qapprox}, this statement is also true when $\tau(\zB)$ varies slowly.

Next, we need to glue the $\{\chi_j\}_{j=0}^{\lvk-1}$ across cuts where the $T^2$ fiber undergoes an $\SL(2,\Z)$ transformation.
We construct the expression
$$
\chi\equiv\sum_{j=0}^{\lvk-1}\chi_j(\zB,\bzB)\Theta_{j,\lvk}(\zF,\tau(\zB)),
$$
where the theta function $\Theta_{j,\lvk}(u,\tau)$ was defined in \eqref{eqn:Thetafn}, and naturally corresponds to the ground state $\ket{j}$. We require the form $\chi$ to be single-valued across cuts, which follows from the fact that the collection of $\Theta$-functions transforms in a dual way to $\chi_j$, since the $\Theta$-functions transform like the states $\ket{j}$.
The $\Theta$-functions are holomorphic and so $\chi$ is in the $d\bzB\bpartial_\bzB$-cohomology.
The $(0,0)$-form and $(1,0)$-form in $\chi$ are
$$
\chi^{(0,0)}+\chi^{(1,0)}\equiv
\sum_{j=0}^{\lvk-1}\omega_j\Theta_{j,\lvk}(\zF,\tau(\zB))
+\sum_{j=0}^{\lvk-1}\omega_{j,\zi}\Theta_{j,\lvk}(\zF,\tau(\zB))d\zB,
$$
which is a formal sum of a holomorphic section $\chi^{(0,0)}$ of a certain line-bundle $\Lbd$ (to be described in more detail below) and a holomorphic differential $\chi^{(1,0)}$ with coefficients in $\Lbd$.
The $(0,1)$ and $(1,1)$ components
$$
\chi^{(0,1)}+\chi^{(1,1)}\equiv
\sum_{j=0}^{\lvk-1}\omega_{j,\bzi}\Theta_{j,\lvk}(\zF,\tau(\zB))d\bzB
+\sum_{j=0}^{\lvk-1}\omega_{j,\zi\bzi}\Theta_{j,\lvk}(\zF,\tau(\zB))d\zB\wedge d\bzB
$$
are of course not holomorphic.
They are required to be in the $d\bzB\bpartial_\bzB$ cohomology, but this cohomology is not so convenient to work with.
We will now show that it can be converted to the ordinary $\bpartial$ cohomology.
Consider for example the $(0,1)$ Dolbeault cohomology with coefficients in some line-bundle $\Lbd$ over $K_3$.
Since $\bpartial=d\bzB\bpartial_\bzB+d\bzF\bpartial_\bzF$, we only need to show that every $\bpartial$-cohomology class has a representative that has no terms in it of the form $(\cdots)d\bzF$ --- in other words, its restriction to any elliptic fiber vanishes.
Let $\varphi$ be a $\bpartial$-closed $(0,1)$ form with coefficients in $\Lbd$ over $K_3$. Let $c_1(\Lbd)$ be the first Chern class.
The restriction of $c_1(\Lbd)$ to any elliptic fiber is $\lvk$ times the generator of $H^{(1,1)}(T^2,\Z)$, and since $\lvk$ is a positive integer, Serre duality implies that $H^1(T^2,\Lbd\lvert_{T^2})=0$. Now cover the base $\efBase$ with contractible open sets, such that each contains at most one special point $\zB_j$. We thus have $\efBase=\bigcup_\alpha\Patch_\alpha$ written as a union of open patches. Let $\pi^{-1}\Patch_\alpha$ be the pre-image of the patch $\Patch_\alpha$ under the projection map $\pi:K_3\rightarrow\efBase$ (of the elliptic fibration). We have $H^1(\pi^{-1}\Patch_\alpha,\Lbd)=0$ and so the restriction of $\varphi$ to $\pi^{-1}\Patch_\alpha$ can be written as $\bpartial\psi_\alpha$ for some local function $\psi_\alpha$.
(This is also true if $\Patch_\alpha$ contains a $\zB_j$, by an explicit computation.)
On $\Patch_{\alpha\beta}\equiv\Patch_\alpha\bigcap\Patch_\beta$ we find that $\psi_{\alpha\beta}\equiv\psi_\alpha-\psi_\beta$ is holomorphic.
Let $\{\rho_\alpha\}_\alpha$ be a partition of unity for the base $\CP^1\simeq\efBase$ (i.e., $\rho_\alpha(\zB,\bzB)$ is a function with support on $\Patch_\alpha$ and $\sum_\alpha\rho_\alpha = 1$ everywhere). Define $g_\alpha\equiv\sum_\beta\rho_\beta\psi_{\alpha\beta}$.
Then the functions $f_\alpha\equiv\psi_\alpha-g_\alpha$ patch together to a global function on $K_3$ which we denote by $f.$ (The last statement follows because, as is easy to see, $f_\alpha=f_\beta$ on $\Patch_{\alpha\beta}.$)
It is then easy to check that the restriction of $\varphi-\bpartial f$ to any elliptic fiber is zero. This completes the proof and demonstrates that we can work with the standard $\bpartial$ differential operator instead of the cumbersome $d\bzB\bpartial_\bzB.$

Let $\CanBdl_\efBase$ be the canonical bundle of the base $\efBase\simeq\CP^1$ (whose transition functions are the Jacobians $d\zB_\alpha/d\zB_\beta$) and let $\CanBdl'\equiv\pi^*\CanBdl_\efBase$ be the pullback of $\CanBdl_\efBase$ under the projection map $\pi: K_3\rightarrow\efBase$. We can now summarize:
\begin{itemize}
\item
Ground states of the form
$\omega_j(\zB,\bzB)\ket{j}$ correspond to elements of $H^0(K_3,\Lbd)$;
\item
Ground states of the form $\omega_{j,\zi}(\zB,\bzB)\bpsiMode^\zi_0\ket{j}$ correspond to elements of $H^0(K_3,\Lbd\otimes\CanBdl')$;
\item
Ground states of the form
$\omega_{j,\bzi}(\zB,\bzB)\psiMode^\bzi_0\ket{j}$
correspond to elements of $H^1(K_3,\Lbd)$;
\item
Ground states of the form
$\omega_{j,\zi\bzi}(\zB,\bzB)\bpsiMode^\zi_0\psiMode^\bzi_0\ket{j}$
correspond to elements of $H^1(K_3,\Lbd\otimes\CanBdl')$;
\end{itemize}
Let us now be more specific about $\Lbd$.
Let $\omega_\efBase$ be the $(1,1)$-cohomology class that is Poincar\'e dual to the zero section $\efBase_0$ ($x=y=\infty$) of the elliptic fibration, whose homology class we denote by $[\efBase_0].$ Let $[\efFiber]$ denote the homology class of the fiber and let $\omega_\efFiber$ be the $(1,1)$-cohomology class that is Poincar\'e dual to $[\efFiber]$.
The intersection numbers are
\be\label{eqn:efIntersection}
[\efFiber]\cdot[\efFiber] = 0\,,\qquad
[\efBase_0]\cdot[\efBase_0] = -2\,,\qquad
[\efBase_0]\cdot[\efFiber] = 1\,.
\ee
We have 
\be\label{eqn:c1Can}
c_1(\CanBdl')=-2\omega_\efFiber\,,\qquad
c_1(\Lbd)=\lvk\omega_\efBase + n\omega_\efFiber\,,
\ee
where $n\in\Z$ still needs to be determined. We determined the coefficient $\lvk$ of $\omega_\efBase$ by the requirement that, when restricted to a fiber, $c_1(\Lbd)$ should be $\lvk$ times the generator of the second cohomology of the fiber.

To proceed, we note that the Hirzebruch-Riemann-Roch theorem for $K_3$-surfaces states that
\be\label{eqn:HRRL}
h^0(K_3,\Lbd)-h^1(K_3,\Lbd) = 
h^0(K_3,\Lbd)-h^1(K_3,\Lbd)+h^2(K_3,\Lbd)=2+\tfrac{1}{2}c_1(\Lbd)^2=2+\lvk n -\lvk^2\,,
\ee
where we used the vanishing of  $h^2(K_3,\Lbd)$, which follows from Serre duality [$h^2(K_3,\Lbd)=h^0(K_3,\Lbd^{-1})$] and the fact that $\Lbd^{-1}$ has no holomorphic sections since it restricts to a line bundle with a negative first Chern class on an elliptic fiber.
Similarly,
\be\label{eqn:HRRLK}
h^0(K_3,\Lbd\otimes\CanBdl')-h^1(K_3,\Lbd\otimes\CanBdl') = 
2+\tfrac{1}{2}c_1(\Lbd\otimes\CanBdl')^2
=2+\lvk (n-2) -\lvk^2\,,
\ee
We can now calculate the total contribution of these ``perturbative states'' to the Witten index:
\be\label{eqn:Ipert}
I_{\text{pert}} = -\left\lbrack
h^0(K_3,\Lbd)-h^1(K_3,\Lbd)-h^0(K_3,\Lbd\otimes\CanBdl')+h^1(K_3,\Lbd\otimes\CanBdl')
\right\rbrack
= -2\lvk = -4.
\ee
The overall $(-)$ sign comes because, as we have seen in \secref{subsec:F}, the ground state $\ket{j}$ has odd fermion number $(-)^F=-1$.

The result \eqref{eqn:Ipert} is not sensitive to the value of $n$, but for completeness let us determine $n$ and the individual dimensions of the cohomologies.
The dimension $h^0(K_3,\Lbd)$ of the space of holomorphic sections of a line bundle $\Lbd$ over $K_3$ can be determined as follows (see \cite{Friedman:1997yq} for the relevant mathematical background).
The question is equivalent to asking for the dimension of the space of meromorphic functions whose divisor of poles is no bigger than $n\efFiber + \lvk\efBase_0$ (i.e., it has a pole of order no bigger than $\lvk$ on $\efBase_0$ and of order no bigger than $n$ on $\efFiber$). For $\lvk=2$, the general function with this property is
$$
f(z,x) = P_n(z) + Q_{n-4}(z) x,
$$
where $P_n$ is a polynomial of degree $n$ in $z$ and $Q_{n-4}$ is a polynomial of degree $(n-4)$. This is because in terms of $\zF$, $x$ is proportional to the Weierstrass function $\wp(\zF;\tau)$, which has a pole of order $2$ at $\zF=0$. Moreover, near $z=\infty$ the elliptic fibration equation \eqref{eqn:efxy} makes sense in coordinates $z'=1/z$ with $x'=x/z^4$ and $y'=y/z^6$. Therefore, near $z'=0$ we find that $f(1/z',x'/{z'}^4)$ has a pole of order $n$ in $z'$. We can therefore calculate the dimension of $h^0(K_3,\Lbd)$ as the number of independent coefficients in $f(z,x)$ and find:
\be\label{eqn:h0k=2}
(\lvk=2)\qquad
h^0(K_3,\Lbd) = \begin{cases}
2n-2 & \text{for $n\ge 3$} \\
n+1 & \text{for $0\le n\le 2$} \\
0 & \text{for $n<0$} \\
\end{cases}
\ee
For $\lvk=3$ a similar argument shows that 
$$
P_n(z) + Q_{n-4}(z)x + R_{n-6}(z)y
$$
is the general solution, so that
\be\label{eqn:h0k=3}
(\lvk=3)\qquad
h^0(K_3,\Lbd) = \begin{cases}
3n-7 & \text{for $n\ge 5$} \\
2n-2 & \text{for $3\le n\le 4$} \\
n+1 & \text{for $0\le n\le 2$} \\
0 & \text{for $n<0$} \\
\end{cases}
\ee
while for $\lvk=1$ we are left with only a polynomial $P_n(z)$ with
\be\label{eqn:h0k=1}
(\lvk=1)\qquad
h^0(K_3,\Lbd) = \begin{cases}
n+1 & \text{for $0\le n$} \\
0 & \text{for $n<0$} \\
\end{cases}
\ee
We can now calculate $h^1(K_3,\Lbd)$ from \eqref{eqn:HRRL}, and in particular we find that
\be\label{eqn:h1}
h^1(K_3,\Lbd) = 0\quad\text{for}\quad\begin{cases}
n\ge 0,\qquad\lvk=1\\
n\ge 3,\qquad\lvk=2\\
n\ge 5,\qquad\lvk=3\\
\end{cases}
\ee
For $h^0(K_3,\Lbd\otimes\CanBdl')$ and $h^1(K_3,\Lbd\otimes\CanBdl')$ we get similar results after replacing $n\rightarrow n-2$ in \eqref{eqn:h0k=2}-\eqref{eqn:h1}.

Now let us determine $n$.
The first Chern class of a line bundle can be calculated if a norm on the $\C$-fibers of the line bundle is given (with the assumption that the norm is independent of the patch used to calculate it). Thus, if $s$ is a local section of $\Lbd$ then
(see for instance \cite{GriffithsHarris} p148):
$$
c_1(\Lbd) = -\frac{i}{2\pi}[\partial\bpartial\log(\|s\|^2)]\,,
$$
where $\|s\|$ is the norm of $s$ (as a function of the point on $K_3$) and $[(\cdots)]$ is the cohomology class of the $2$-form $(\cdots)$.
In our case a natural norm is given as follows.
Consider for example the $0$-forms 
$$
\chi^{(0,0)}\equiv\sum_{j=0}^{\lvk-1}\chi_j^{(0,0)}(\zB,\bzB)\Theta_{j,\lvk}(\zF,\tau(\zB)).
$$
Since, we can write the normalized wave-function on $T^2$ as \eqref{eqn:Lwf}
with $\wfF_{j,\lvk}$ normalized so that $\int| \wfF_{j,\lvk}|^2 d\xf_1 d\xf_2=1$,
we find the total norm 
$$
\|\chi^{(0,0)}\|^2 = \frac{1}{4\pi^2}\int 
(2\lvk\tau_2)^{1/2}e^{-\frac{\lvk\tau_2\xf_2^2}{2\pi}}
|\chi^{(0,0)}|^2
|g_{\zi\bzi}| d^2\zi d\xf_1 d\xf_2
$$
and we can set
$$
\|s\|^2\rightarrow 
\frac{1}{4\pi^2}
(2\lvk\tau_2)^{1/2}e^{-\frac{\lvk\tau_2\xf_2^2}{2\pi}}
|\chi^{(0,0)}|^2\,.
$$
Since $\chi^{(0,0)}$ is holomorphic, and setting $\xf_2 = 2\pi(\zF-\bzF)/(\tau-\btau)$, we find
\be\label{eqn:c1Lbd}
c_1(\Lbd)=
-\frac{i}{2\pi}
\bigl[\partial\bpartial\bigl(\frac{-\lvk\tau_2\xf_2^2}{2\pi}\bigr)\bigr]
-\tfrac{i}{4\pi}\bigl[\partial\bpartial\log\tau_2\bigr]
=
\frac{1}{4\pi^2}\lvk d\xf_1\wedge d\xf_2
+\frac{i}{16\pi\tau_2^2}d\tau\wedge d\btau
\ee
The integral of the second term over a fundamental domain $\FD$ of the $\SL(2,\Z)$ action on the upper-half $\tau$-plane gives
$$
\frac{i}{2\pi}\int_\FD\frac{d\tau\wedge d\btau}{8\tau_2^2} 
= \frac{1}{24}\,.
$$
Since the elliptic fibration equation \eqref{eqn:efxy} describes the base $\efBase$ (given by the locus of $\xf_1=\xf_2=0$) as a $24$-fold cover of the fundamental domain $\FD$, we find that the pullback of $c_1(\Lbd)$ to $\efBase$ is $24\times (\tfrac{1}{24})=1$.
Thus
$$
c_1(\Lbd) =\lvk\omega_\efBase + (1+2\lvk)\omega_\efFiber\,.
$$
[Note that $\int_{\efBase_0}\omega_\efBase = [\efBase_0]\cdot[\efBase_0]=-2$, so $\int_{\efBase_0} c_1(\Lbd) = (1+2\lvk-2\lvk)$.]
In particular, \eqref{eqn:h1} implies that 
$$
h^1(K_3,\Lbd)=h^1(K_3,\Lbd\otimes\CanBdl')=0\,,\qquad
\lvk=1,2,3,
$$ 
and all the states can be described by holomorphic sections of either $\Lbd$ or $\Lbd\otimes\CanBdl'$. For $\lvk=2$, for example, we find
$$
h^0(K_3,\Lbd)=8\,,\qquad
h^0(K_3,\Lbd\otimes\CanBdl')=4\,.
$$

\subsection{The Witten Index}
\label{subsec:WittenIndexFM}

Mirror-symmetry of $K_3$ can formally be described \cite{Morrison:1996ac,Bartocci:1997cd} by its action on what is known as a {\it Fourier-Mukai vector.} This is an element $v\equiv v_0+v_2+v_4$ in the lattice $H^0(K_3,\Z)\oplus H^2(K_3,\Z)\oplus H^4(K_3,\Z)$, and we define $v^2\equiv 2v_0\cdot v_4 + v_2^2$ via the standard intersection form on cohomology.
A mirror-symmetry element is described by a linear transformation on this vector $v$ that preserves the lattice $H^0(K_3,\Z)\oplus H^2(K_3,\Z)\oplus H^4(K_3,\Z)$ and preserves $v^2.$ 
Physically, in the context of type-IIA string theory compactification on $K_3\times\R^{5,1}$, we understand $v$ as a vector of D-brane charges (D$0$, D$2$, and D$4$). Mirror symmetry converts a combination of pointlike D$0$'s,  D$2$'s that wrap (real) surfaces in $K_3$, and D$4$'s that wrap the entire $K_3$ into a similar combination but with different charges. The advantage of describing a duality in terms of a Fourier-Mukai vector is that the duality naturally corresponds to an $O(20,4,\Z)$ matrix.

We can now calculate the Witten index of the theory with the mirror-symmetry twist by requiring modular invariance of the torus partition function. Explicitly, we insert the mirror-symmetry operator $\OpM$ at $\sigma^1=0$, and we compactify (Euclidean) time $\sigma^0$ on $S^1$ with periodic boundary conditions for the fermions.
We need to calculate the partition function, which we do by interchanging the roles of $\sigma^0$ and $\sigma^1$. If $\sigma^0$ plays the role of ``space'', the Hilbert space is the space of Ramond-Ramond ground states, which is naturally mapped to the Fourier-Mukai vector space.\footnote{In the string theory context each D-brane charge can be mapped to a (possibly fractional) set of Ramond-Ramond fluxes, as measured at infinity. The operators that correspond (under the CFT state-operator correspondence) to the Ramond-Ramond ground states of the $\sigma$-model are factors in the vertex operators that correspond to the type-II Ramond-Ramond fluxes.}
Letting $\mvM$ denote the $O(20,4,\Z)$ matrix that describes the action of $\OpM$ on the Fourier-Mukai vector, we calculate the Witten index as
\be\label{eqn:ImvM}
I(\mvM) = \tr\{(-1)^F\mvM\} = \tr\mvM.
\ee
We have set $(-)^F=1$, since we only have even-dimensional cohomology.

In the complex structure that is compatible with the elliptic fibration (i.e., $\pi: K_3\rightarrow\efBase$ is holomorphic), the nonzero Hodge numbers are $h^{0,0}=h^{2,2}=h^{2,0}=h^{0,2}=1$ and $h^{1,1}=20.$
Let us now review some facts about the cohomology group $H^2(K_3,\Z)$. We can construct a basis of the Poincar\'e dual homology group $H_2(K_3,\Z)$ as follows. First, let $[\efFiber]$ and $[\efBase_0]$ be the homology classes defined above equation \eqref{eqn:efIntersection}, which correspond to the holomorphic submanifolds given by $\{\zB=0\}$ and $\{x=y=\infty\}$, respectively. 
The remaining $18$ independent generators of $H^2(K_3,\Z)$ cannot be represented by analytic submanifolds but can be constructed as follows.
Let $\gamma$ be a smooth path on $\efBase$ from one of the special points $\zB_i$ to another $\zB_j$ ($i\neq j$), avoiding all other special points. While $\tau(\zB)$ is a multivalued function, undergoing $\SL(2,\Z)$ transformations along cuts,
we can find a small neighborhood of $\gamma\setminus\{\zB_i,\zB_j\}$ on which $\tau(\zB)$ is represented by an analytic function without cuts.
If in that representation $\tau(\zB_i)=\tau(\zB_j)=\infty$ (in general, the limits $\tau(\zB_i)$ and $\tau(\zB_j)$ are only guaranteed to be $\SL(2,\Z)$ equivalent) then we can construct a $2$-cycle in $K_3$ by taking the basic $\alpha$ $1$-cycle of $\efFiber$ (from $\xf_1=0$ to $\xf_1=2\pi$ at a fixed $\xf_2$), and dragging it along $\gamma.$ This cycle will shrink to a point at both ends of $\gamma$ and hence form a closed $2$-cycle.
It can be shown that in homology there are $18$ linearly independent $2$-cycles of this type, with intersection form that is equivalent to the $E_8\oplus E_8\oplus \mathbb{H}$ lattice, where $\mathbb{H}$ is the $2$-dimensional lattice $\Z^2$ with intersection form $(n_1,n_2)^2=2 n_1 n_2$ [for $(n_1,n_2)\in\Z^2$].

Let $\omega_\efFiber,\omega_\efBase$ be the Poincar\'e dual $(1,1)$ cohomology classes such that
$$
\int_{[\efFiber]}\omega_\efFiber=0\,,\qquad
\int_{[\efBase]}\omega_\efFiber=\int_{[\efFiber]}\omega_\efBase=1\,,\qquad
\int_{[\efBase]}\omega_\efBase=-2\,.
$$
For generic $f_8$ and $g_{12}$, all the integral $(1,1)$ forms are linear combinations of $\omega_\efFiber$ and $\omega_\efBase$ with integer coefficients.
Let $\vertBF\subset H^2(K_3,\Z)$ be the sublattice of elements that are orthogonal (in terms of the intersection pairing) to both $\omega_\efFiber$ and $\omega_\efBase.$ It is generated by the cohomology classes of $2$-cycles constructed from the paths between $\zB_i$ and $\zB_j$ as described above.
We can now describe $\mvM$. 
Let $x_0$ and $x_4$ be generators of $H^0(K_3,\Z)$ and $H^4(K_3,\Z)$ so that $x_0\cdot x_4=1$.
Then $\mvM$ acts as $(+1)$ on $\vertBF$ and acts as follows on the remaining generators:
$$
\mvM(x_0)=\omega_\efFiber\,,\qquad
\mvM(x_4)=\omega_\efBase+\omega_\efFiber\,,\qquad
\mvM(\omega_\efFiber)= -x_0\,,\qquad
\mvM(\omega_\efBase)= x_0-x_4\,.
$$
According to the prescription of \secref{subsec:SUSY}, this transformation preserves two pairs of left and right moving supersymmetries. 
We note that for $(2,2)$ models, mirror symmetry always preserves all the left-moving supercharges and only one combination of the right-moving supercharges (see, e.g., \cite{Hori:2000kt}). Since we are dealing with a $(4,4)$ model, there are two additional left and right moving supercharges that are not part of the $(2,2)$ superconformal algebra, and one combination of the two additional right-moving supercharges is also preserved by $\mvM$.
Calculation of the Witten index now gives:
\be\label{eqn:ImvMcalc}
I(\mvM) = \tr\{(-1)^F\mvM\} = 20.
\ee
In \secref{subsec:PertAnal} we calculated the contribution of $I_{\text{pert}}=-4$ to the Witten index from states with extended wave-functions along $\efBase$.
Comparing \eqref{eqn:Ipert} to \eqref{eqn:ImvMcalc}, we see that we need to account for an additional contribution of $24$. Since there are exactly $24$ special points of $\efBase$ with singular fibers, it stands to reason that there are extra ``bound states'' localized near these special points. We assume that the elliptic fibration is generic, so that the $\SL(2,\Z)$ monodromies (in $\tau$) around the special points are all conjugate to $\tau\rightarrow\tau+1$, and the behaviors of the fibration near all the special points are equivalent. Thus, if there is a contribution to the Witten index from localized ground states, it should be a multiple of $24$. It is indeed pleasing that our analysis yielded a discrepancy of exactly $24$.
In the next subsection we will bring supporting evidence for the conjecture that each special point contributes one bound state.

This result can be extended to other values of $\lvk$ (namely $\lvk=1$ or $\lvk=3$) as follows.
We work at the value $\rho=e^{i\pi/3}$ (where $\rho$ is the complexified area of the fiber).
This value is fixed by $O(20,4,\Z)$ transformations that act as in \secref{subsec:OtherSL} on the fiber.
We can construct them by compounding $\mvM$ with an operator $\mvT$ that acts as $(+1)$ on $\vertBF$ and acts as follows on the remaining generators:
$$
\mvT(x_0)=x_0\,,\qquad
\mvT(x_4)=x_4+\omega_\efBase + x_0\,,\qquad
\mvT(\omega_\efFiber)= \omega_\efFiber -x_0\,,\qquad
\mvT(\omega_\efBase)= \omega_\efBase + 2x_0\,.
$$
This is the transformation that shifts the $B$-field by $\omega_\efBase$.
We find,
$$
I(\mvM\mvT) = \tr\{(-1)^F\mvM\mvT\} = 18\,,\qquad
I(\mvM\mvT^{-1}) = \tr\{(-1)^F\mvM\mvT^{-1}\} = 22\,.
$$
Locally on the base, $\mvM\mvT$ and $\mvM\mvT^{-1}$ act as $\SL(2,\Z)$ duality transformations of the fiber with $\lvk=3$ and $\lvk=1$ respectively. The index can therefore be written generally as $I=24-2\lvk$ (for $\lvk=1,2,3$) and the contribution of $-2\lvk$ is explained as in \eqref{eqn:Ipert}.

\subsection{The singular fibers}
\label{subsec:SF}

We can gain more insight about the behavior near a singular fiber by embedding our problem into string theory. A natural string theory setting is a compactification of type-IIA string theory (it is more convenient to use type-IIA for reasons to become clear shortly) on a $K_3$ at a point in moduli space that is invariant under the mirror-symmetry element $\OpM$ considered in \secref{subsec:WittenIndexFM}. We then compactify one of the remaining directions, say $x_5$, on $S^1$ of radius $R$ and insert an $\OpM$-twist at $x_5=0$.
In order to preserve some amount of target-space supersymmetry, we also insert at that point a suitable rotation $\gamma\in\Spin(4)$ in directions $6,\dots,9$.\footnote{The amount of target space supersymmetry is determined as follows. 
Similarly to \secref{subsec:SUSY}, set $K\equiv O(20)\times SO(4)$ and let $gK\in O^{+}(20,4,\R)/K$ be the point in moduli space, represented as a coset of groups. If $\OpM\in O^{+}(20,4,\Z)$ preserves $gK$ then $g^{-1}\OpM g\in K.$ Let $\rho\in SO(4)$ be the projection of $g^{-1}\OpM g$ to the $SO(4)$ factor of $K$. To construct the background we have to specify the action of $\OpM$ on target-space fermions, which requires a lift of $\rho$ to the spin group. Let $e^{\pm i\varphi_1}$ and $e^{\pm i\varphi_2}$ be the eigenvalues of (the lift of) $\rho$ in the spinor representation $(2,1)\oplus (1,2)$. Then the action of $\OpM$ on the supercharges has eigenvalues $e^{\pm i\varphi_j}$ ($j=1,2$), and the number of preserved SUSY generators is the number of eigenvalues that $\gamma$ has from the set $\{e^{\mp i\varphi_j}\}_{j=1}^2$, when $\gamma$ is expressed in the spinor representation $(2,1)\oplus (1,2)$. This can be contrasted with the worldsheet SUSY discussed in \secref{subsec:SUSY}, where the scalar and tensor representations $2(1,1)\oplus(3,1)\oplus(1,3)$ of $SO(4)$ were used instead of the spinor.} We then look for the ground states of a string state with winding number $w=1$ along direction $x_5$. The twist $\gamma$ can be chosen so as to ensure that the ground states are normalizable and localized at $x_6=x_7=x_8=x_9=0$. 
The worldsheet theory that solves this problem (at string coupling constant zero) is an asymmetric $\Z_4$ orbifold of an $S^1\times K_3$ compactification with $S^1$ of radius $4 R$. Since directions $5,\dots,9$ are described by a flat target space their contribution is easy to analyze and we find that the Ramond-Ramond ground states of the string are in one-to-one correspondence with the states of the $K_3$ CFT compactified on $S^1$ with an $\OpM$-twist.


Now let us focus on the vicinity of a special point $\zB_j.$
{}From the point $\zB_j$ there emanates a cut that is accompanied by the $\SL(2,\Z)$ transformation $\tau\rightarrow\tau+1$. 
We will now perform a series of dualities to relate our problem to a problem of quantizing an $\Omega$-deformed \cite{Nekrasov:2002qd,Nekrasov:2010ka} instanton moduli space. Let $\xf_1,\xf_2$ be coordinates on the fiber $\efFiber$ over a certain point $\zB$ of the base such that direction $\xf_1$ corresponds to the cycle that shrinks to zero at $\zB_j$. Now perform a fiberwise T-duality on direction $\xf_1$. This converts the fiber to a $T^2$ with complex structure $\wtau=i$ and area $\tau_2$ that becomes infinite at $\zB_j$. The monodromy $\tau\rightarrow\tau+1$ around the cut shows that in this T-dual type-IIB background we have an NS$5$-brane at $\zB_j$ that extends in directions $5,\dots,9$, and is localized at the dual fiber location $(\wxf_1,\xf_2)=(0,0)$. The mirror-symmetry twist becomes a geometrical twist that acts as a rotation by $\pi/2$ in directions $\wxf_1,\xf_2$. There are two fixed points
\be\label{eqn:wxfxf}
(\wxf_1,\xf_2)=\text{$(0,0)$ or $(\pi,\pi)$.}
\ee
The $(\pi,\pi)$ fixed point is far from the NS$5$-brane and so is not expected to have any additional bound states.
Let us now analyze the $(0,0)$ point by replacing the dual fiber with an $\R^2$.
We now have a fundamental string wrapped around direction $x_5$ with a geometrical twist in directions $\wxf_1,\xf_2,x_6,x_7,x_8,x_9$ and in the vicinity of an NS$5$-brane spanning directions $x_5,\dots,x_9$ and localized at $\zB=\zB_j$ and $\wxf_1=\xf_2=0$. The question is whether we get additional states localized at $\zB=\zB_j$. Such states will be related to the existence of bound states of the string with the NS$5$-brane. We can generalize the question to one about bound states of a string with $n$ NS$5$-branes, which corresponds to a monodromy of $\tau\rightarrow\tau+n$ in the elliptic fibration as would appear when $n$ special points coincide (forming a singular fiber of type $A_{n-1}$).
We can solve the question by performing S-duality, converting the string and NS$5$-branes into a D$1$ and $n$ D$5$-branes. (The geometrical twists are similar to those used in the $\Omega$-deformation context to regularize the moduli space of instantons \cite{Nekrasov:2002qd,Nekrasov:2010ka}.) We can also regularize the moduli space by adding NSNS $B$-field components $B_{67}$ and $B_{89}$ to turn the moduli space of instantons into instantons on a noncommutative space \cite{Nekrasov:1998ss,Berkooz:1998st}.
Let us now be more specific about the rotation $\gamma$ that was introduced in order to preserve SUSY and eliminate zero modes. We can pick it to be a simultaneous rotation by $-\pi/4$ in two planes, the plane $x_6-x_7$ and the plane $x_8-x_9$, so that combined with the $\pi/2$ rotation in the plane of $\wxf_1-\xf_2$ the background will preserve $1/8$ (target space) supersymmetry. At weak string coupling-constant and large $R$, the D$1$-D$5$ dynamics is described locally \cite{Witten:1995gx,Douglas:1995bn} by a point on the moduli space of $U(n)$ instantons (on a noncommutative $\R^4$ in directions $6,\dots,9$) that varies as a function of $x_5$, and with a twist at $x_5=0$ that corresponds to a rotation by $(-\pi/4,-\pi/4)$ in the planes $6-7$ and $8-9$, and a rotation by $\pi/2$ in the plane $1-2$ (which acts only on the fermionic degrees of freedom). The ground states are found by counting the number of instanton solutions that are invariant under the $(-\pi/4,-\pi/4)$ rotation. Consider, for example, the case of an $SU(2)$ instanton and start with a commutative $\R^4$. A rotation in $\R^4$ can be compensated by a global $SU(2)$ gauge transformation as long as the center of the instanton is at the origin. However, we are forbidding such transformations that modify the behavior at infinity (in directions $6,\dots,9$) and so we are left with only the zero-size instantons as a solution. These need to be regularized by turning $\R^4$ into a noncommutative $\R^4$, as mentioned above.

Thus, in order to proceed we recall the extension \cite{Nekrasov:1998ss,Berkooz:1998st} of the ADHM construction for $U(n)$ instantons at level $m$ (for us $m=1$) on a noncommutative $\R^4$. We take complex coordinates $(z_0, z_1)$ on $\R^4\simeq\C^2$ with noncommutativity given by
$$
[z_0,z_0^\dagger]=[z_1,z_1^\dagger]= -\frac{\NCzeta}{2}\,,
$$
where $\NCzeta$ is a positive constant. One then picks $n\times n$ matrices $B_0, B_1$, an $n\times m$ matrix $X$, and an $m\times n$ matrix $Y$ (all matrices have constant complex entries) that satisfy the quadratic algebraic relations:
\be\label{eqn:ADHM}
[B_0, B_1]+ X Y = 0,\qquad
[B_0, B_0^\dagger]+[B_1, B_1^\dagger]+X X^\dagger - Y^\dagger Y = \NCzeta\Id_{n\times n}\,,
\ee
where $\Id_{n\times n}$ is the identity matrix. The moduli space of solutions is then the solution to \eqref{eqn:ADHM} subject to a $U(m)$ gauge equivalence defined to act on $X$ and $Y$ as 
\be\label{eqn:XYLambda}
X\rightarrow X\Lambda^{-1}\,,\qquad
Y\rightarrow\Lambda Y\,,\qquad\Lambda\in U(m).
\ee
One then looks for a $(2n+m)\times n$ matrix solution $\psi(z_0,z_1)$ for the linear equations
$$
\begin{pmatrix}
B_0 - z_0 &\quad& z_1-B_1 &\quad& X \\
B_1^\dagger - z_1^\dagger &\quad& B_0^\dagger -z_0^\dagger &\quad& Y^\dagger \\
\end{pmatrix}\psi = 0
$$
where $z_0, z_1$ are short for $z_0\Id_{n\times n}, z_1\Id_{n\times n}$.  One then constructs the gauge field as $A_\mu=\psi^\dagger\partial_\mu\psi$. We need an instanton solution that is invariant under the rotation
$$
(z_0,z_1)\mapsto (e^{i\pi/4}z_0, e^{i\pi/4}z_1).
$$
This rotation acts on the ADHM fields as
$$
B_0\rightarrow e^{\frac{i\pi}{4}}B_0\,,\qquad
B_1\rightarrow e^{\frac{i\pi}{4}}B_1\,,\qquad
X\rightarrow e^{\frac{i\pi}{4}}X\,,\qquad
Y\rightarrow e^{\frac{i\pi}{4}}Y\,.
$$
and it is easy to see that the only points in moduli space that are invariant under this rotation [up to a gauge transformation \eqref{eqn:XYLambda}] are those with $B_0=B_1=0$ and $Y=0$.
For $m=1$, it is easy to see that the surviving moduli space is $\CP^{n-1}$.
The Hilbert space of localized states coming from an $A_{n-1}$ singular fiber is therefore equivalent to the cohomology of $\CP^{n-1}$. In particular, a generic singular fiber ($A_0$ type) carries one extra state.

\section{Other examples of mirror twists in the orbifold limit}
\label{sec:OtherX}

In the second part of this paper we will explore the ground states of theories with twists by other elements of the duality group $SO(20,4,\Z)$ and their possible relation to geometric quantization of $K_3$. We will perform the computations in the orbifold limit $T^4/\Z_2$.
We start by reviewing some known properties of these orbifolds (see \cite{Aspinwall:1996mn} for a comprehensive review).

\subsection{$K_3$ as a $T^4/\Z_2$ orbifold}
\label{subsec:CohoK3}

We take $(\xf_1,\xf_2,\xf_3,\xf_4)$ as coordinates on $T^4$, where each $\xf_i$ ($i=1,\dots,4$) has periodicity $2\pi.$ We also define the complex structure by
$$
z = \frac{1}{2\pi}(\xf_1 +\tau\xf_2)\,,\qquad
z' = \frac{1}{2\pi}(\xf_3 +\tau'\xf_4)\,.
$$
The $\Z_2$ orbifold is defined as
\be\label{eqn:zzmzz}
(z,z')\simeq (-z,-z')\,.
\ee
This orbifold has $16$ fixed points
$$
(\xf_1,\xf_2,\xf_3,\xf_4) 
= (\epsilon_1\pi,\epsilon_2\pi, \epsilon_3\pi,\epsilon_4\pi);
\qquad
(\epsilon_1,\dots,\epsilon_4\in\{0,1\})\,,
$$
which we denote by $P(\epsilon_1,\dots,\epsilon_4).$
A smooth complex manifold can be constructed from this orbifold by ``blowing-up'' these fixed points and replacing each $P(\epsilon_1,\dots,\epsilon_4)$ with a $\CP^1$ that we denote by $E(\epsilon_1,\dots,\epsilon_4).$ We recall that blowing up the origin of $\C^2/\Z_2$ is done by replacing the singular orbifold $(z,z')\simeq(-z,-z')$ with the subspace of $\C^2\times\CP^1$ described by coordinates $(z,z',[s,t])$ (where we use $(z,z')$ as coordinates on $\C^2$ and $[s,t]\simeq [\lambda s,\lambda t]$ as homogeneous coordinates on $\CP^1$) that satisfy $zt=z's$, so that for $(z,z')\neq(0,0)$ there is a unique solution $[s,t]=[z,z']$ but for $z=z'=0$ the entire $\CP^1$ survives. Blowing up an orbifold point in $T^4/\Z_2$ is done by replacing a small neighborhood of that point with a neighborhood of the blown-up origin of $\C^2/\Z_2.$ So far we only discussed the complex structure. The $\sigma$-model requires also a complexified K\"ahler metric which is given by a K\"ahler $2$-form $k$ and NSNS B-field $B.$
It was shown in \cite{Aspinwall:1995zi} that the orbifold limit of the $\sigma$-model corresponds to the limit where the area of each $E(\epsilon_1,\dots,\epsilon_4)$ vanishes, while the flux of $B$ on each $E(\epsilon_1,\dots,\epsilon_4)$ remains $\pi.$
This orbifold has a $\Z_2^4$ group of isometries. The element $\Zivp(\varpi_1,\dots,\varpi_4)$ that is labeled by $\varpi_1,\dots,\varpi_4\in\Z_2$ (which we treat as numbers modulo $2$) acts as
\be\label{eqn:Ziv}
\Zivp(a_1,\dots,a_4): 
(\xf_1,\xf_2,\xf_3,\xf_4)\mapsto
(\xf_1+\pi\varpi_1,\xf_2+\pi\varpi_2,\xf_3+\pi\varpi_3,\xf_4+\pi\varpi_4)\,,\qquad
a_1,\dots,a_4\in\{0,1\}.
\ee
Each of these isometries has fixed points. For example $\Zivp(1,1,1,1)$ has fixed points when each coordinate $\xf_1,\dots,\xf_4$ is $\pm\pi/4$, and with the identification \eqref{eqn:zzmzz}, this gives $8$ fixed points. Other than the identity, these isometries act nontrivially on the cycles $E(\epsilon_1,\dots,\epsilon_4)$:
$$
E(\epsilon_1,\dots,\epsilon_4)\xrightarrow{\Zivp(a_1,\dots,a_4)}
E(\epsilon_1+a_1,\dots,\epsilon_4+a_4)\,,
$$
where $(\epsilon_i+a_i)$ is understood modulo $2$.


We will need a convenient basis for the cohomology $H^2(K_3,\Z)$.
Let us start by defining the following elements of $H^2(K_3,\Z)$: 
\begin{itemize}
\item
$[E(\epsilon_1,\dots,\epsilon_4)]$ (with $\epsilon_i\in\{0,1\}$) denotes the Poincar\'e dual of one of the $16$ $\CP^1$ submanifolds $E(\epsilon_1,\dots,\epsilon_4)$ at the intersection of $z=\frac{1}{2}(\epsilon_1+\tau\epsilon_2)$ and $z'=\frac{1}{2}(\epsilon_3+\tau'\epsilon_4)$;
\item
$[M(i,j)]=-[M(j,i)]$ (with $1\le i<j\le 4$) corresponds to the Poincar\'e dual of the submanifold  given by the two equations $\xf_i=\pm c_i$ and $\xf_j=\pm c_j$ (with the $\pm$ signs correlated), for some generic constants $c_i, c_j$ (not equal to $0$ or $\pi$). We take the orientation so that the dual class is represented by $\delta(\xf_i-c_i)\delta(\xf_j-c_j)d\xf_i\wedge d\xf_j$;
\end{itemize}
The intersection numbers are:
\be\label{eqn:H2int}
\begin{split}
&[E(\epsilon_1,\dots,\epsilon_4)]\cdot[E(\epsilon_1',\dots,\epsilon_4')]
=-2\delta_{\epsilon_1\epsilon_1'}\delta_{\epsilon_2\epsilon_2'}
\delta_{\epsilon_3\epsilon_3'}\delta_{\epsilon_4\epsilon_4'}
\,,\\
&[E(\epsilon_1,\dots,\epsilon_4)]\cdot[M(i,j)] =0
\,,\qquad
[M(i,j)]\cdot[M(k,l)] =2\epsilon_{ijkl}
\,,
\end{split}
\ee
where $\epsilon_{ijkl}$ is the anti-symmetric Levi-Civita symbol.
In addition to these cohomology classes, $H^2(K_3,\Z)$ also contains some classes that are linear combinations of the $[E(\epsilon_1,\dots,\epsilon_4)]$'s and $[M(i,j)]$'s with fractional coefficients. We define 
$$
[W(i,j,\epsilon_i,\epsilon_j)]
\equiv\tfrac{1}{2}[M(i,j)]-\tfrac{1}{2}\sum_{\epsilon_1'=0}^1\cdots\sum_{\epsilon_4'=0}^1
\delta_{\epsilon_i\epsilon_i'}\delta_{\epsilon_j\epsilon_j'}
[E(\epsilon_1',\dots,\epsilon_4')].
$$
Then, one can show that $[W(i,j,\epsilon_i,\epsilon_j)]\in H^2(K_3,\Z),$ and $H^2(K_3,\Z)$ is spanned by the $[W(i,j,\epsilon_i,\epsilon_j)]$'s and the $[E(\epsilon_1,\dots,\epsilon_4)]$'s (which is an over-complete system).
The $[W(i,j,\epsilon_i,\epsilon_j)]$'s are Poincar\'e dual to homology classes of submanifolds (not necessarily analytic) that can be constructed by setting $c_i$ and $c_j$, in the definition of $[M(i,j)]$ above, to $0$ or $\pi$. For example, $[W(3,4,\epsilon_3,\epsilon_4)]$ is the Poincar\'e dual of the smooth analytic manifold defined by the holomorphic equation,
\be\label{eqn:DefW}
W(3,4,\epsilon_3,\epsilon_4) = \{z'=\tfrac{1}{2}(\epsilon_3+\epsilon_4\tau')\}.
\ee
which has genus $0$.
As another example, $[W(1,3,\epsilon_1,\epsilon_3)]$ can be constructed by starting with the equations $\xf_1=\epsilon_1$ and $\xf_3=\epsilon_3$ for $\epsilon_1,\epsilon_3\in\{0,1\}.$ This manifold is not analytic, of course, and it has a boundary. It intersects $E(\epsilon_1,\epsilon_2',\epsilon_3,\epsilon_4')$ along a circle which divides $E(\epsilon_1,\epsilon_2',\epsilon_3,\epsilon_4')$ into two hemispheres. If we attach one of these hemispheres to the manifold, for each of the four possible combinations of $\epsilon_2'$ and $\epsilon_4'$, we can get a closed manifold whose Poincar\'e dual is given by $[W(1,3,\epsilon_1,\epsilon_3)].$



The equation $z=\pm z_0$, with generic $z_0$, describes a $T^2\subset T^4/\Z_2$. The cohomology class of this $T^2$ is given by
$$
[\efFiber]\equiv[M(1,2)],
$$
and we can view it as the fiber of an elliptic fibration that becomes singular at the $4$ points $z_0=0,\tfrac{1}{2},\tfrac{1}{2}\tau,\tfrac{1}{2}+\tfrac{1}{2}\tau.$
As we take the limit $z_0\rightarrow \tfrac{1}{2}(\epsilon_1+\epsilon_2\tau)$, for example, the fiber $T^2$ turns into a union of the four exceptional divisors $\bigcup\limits_{\epsilon_3,\epsilon_4} E(\epsilon_1,\epsilon_2,\epsilon_3,\epsilon_4)$ ($\epsilon_3,\epsilon_4=0,1$) together with the sphere $W(1,2,\epsilon_1,\epsilon_2)$ counted with multiplicity $2$, so that the cohomology agrees:
$$
2[M(1,2)]=
2[W(1,2,\epsilon_1,\epsilon_2)]
+\sum\limits_{\epsilon_3,\epsilon_4} [E(\epsilon_1,\epsilon_2,\epsilon_3,\epsilon_4)]\,.
$$
The fibration has a section which we take to be $W(3,4,\epsilon_3=0,\epsilon_4=0).$
We define the cohomology class
$$
[\efBase]\equiv [W(3,4,\epsilon_3=0,\epsilon_4=0)]
=\tfrac{1}{2}[M(3,4)]
-\tfrac{1}{2}\sum\limits_{\epsilon_1',\epsilon_2'}
[E(\epsilon_1',\epsilon_2',0,0)].
$$

\subsection{Mirror symmetry}
\label{subsec:MS}
In the orbifold limit the CFT is a free theory.
Consider the space of states of the CFT on $S^1$ with coordinate $0\le\sigma^1\le 2\pi.$
The Minkowski time coordinate is $\sigma^0.$
In the untwisted sector
there are $4$ scalar fields $\scX^\mu$ ($\mu=1,\dots,4$) with oscillator expansions \cite{Green:1987sp,Polchinski:1998rq}:
$$
\scX^\mu(\sigma^0,\sigma^1) = x^\mu +p^\mu\sigma^0
+\frac{i}{\sqrt{2}}
\sum_{\substack{m\in\Z\\m\neq 0}}
\tfrac{1}{m}\left(
   \osc_m^\mu e^{-i m(\sigma^0-\sigma^1)}+\wosc_m^\mu e^{-i m(\sigma^0+\sigma^1)}\right)
$$
as well as four left-moving fermionic $\psi^\mu$ fields and right-moving $\wpsi^\mu$ fields:
$$
\psi^\mu(\sigma^0,\sigma^1)=   
\sum_{m\in\Z}\psi_m^\mu e^{-i m(\sigma^0-\sigma^1)}
\,,\qquad
\wpsi^\mu(\sigma^0,\sigma^1)=   
\sum_{m\in\Z}\wpsi_m^\mu e^{-i m(\sigma^0+\sigma^1)}
\,.
$$
The twisted sector has similar expansions with $m\in\Z+\tfrac{1}{2}$ and with $(x^1,\dots,x^4)$ taking values in one of the $16$ fixed points of the $\Z_2$ action:
$$
(x^1,\dots,x^4) = (\epsilon_1\pi,\dots,\epsilon_4\pi).
$$
In what follows we will only be interested in the ground states.
Thus we let all the oscillators with $m>0$ be in their ground states and we set $p^\mu=0$.
For the untwisted sector this means that we only have to consider the fermionic zero modes. The ground states $\ket{ab}$ are labeled by two (Dirac) spinor indices $a=1,\dots,4$ and $b=1,\dots 4$ on which the zero modes act as:
$$
\psi_0^\mu\ket{ab}={(\gamma^\mu)_a}^{a'}\ket{a'b}\,,\qquad
\wpsi_0^\mu\ket{ab}={(\gamma^\mu)_b}^{b'}\ket{ab'}\,,
$$
where $\gamma^\mu$ are the Dirac matrices (in some arbitrary basis).
The $\Z_2$ orbifold generator acts as $\psi_0^1\cdots\psi_0^4\wpsi_0^1\cdots\wpsi_0^4$ and the surviving $\Z_2$-even states can be written in the form
\be\label{eqn:ketab}
\ket{ab}=(\gamma^0)_{ab}\ket{1234}
+\frac{1}{4}\epsilon_{\mu\nu\sigma\tau}(\gamma^0\gamma^{\mu\nu})_{ab}\ket{\sigma\tau}
+(\gamma^0\gamma^{1234})_{ab}\ket{\varnothing}\,,
\ee
where we have defined a basis of $1+6+1$ $\Z_2$-invariant states $\ket{\varnothing}$, $\ket{\mu\nu}=-\ket{\nu\mu}$, and $\ket{1234}$.
The twisted sector has no fermionic zero modes, and so the ground states are only labeled by the $(x^1,\dots,x^4)$ fixed point as $\ket{\epsilon_1,\dots,\epsilon_4}$.

In string theory these states appear as factors in the worldsheet description of Ramond-Ramond states for type-II compactification on $K_3\times\R^{5,1}$.
In type-IIA, for example, the states that correspond to modes of the RR $0$-form, $2$-form, and $4$-form fields that are massless plane waves in the $\R^{5,1}$ directions and proportional to harmonic $0$-forms, $2$-forms, or $4$-forms along $K_3$ contain $\ket{ab}$ as a factor.
The states of the untwisted sector correspond to harmonic forms whose cohomology class is Poincar\'e dual to a $\Z_2$ invariant form on $T^4$, and the twisted sector states correspond to harmonic forms that are Poincar\'e dual to $E(\epsilon_1,\dots,\epsilon_4).$

{}From this description it is easy to derive the action of several dualities.
We will consider several combinations of the basic operators defined as follows.
We define $\mvS_1$ as the symmetry induced by T-duality on directions $\xf_3,\xf_4$ followed by a rotation $\xf_3\rightarrow\xf_4.$ It acts as  
\be\label{eqn:mvS1}
\begin{split}
&\mvS_1\ket{\varnothing} = \ket{12}
\,,\quad
\mvS_1\ket{12} = -\ket{\varnothing}
\,,\quad
\mvS_1\ket{34} = 2 \ket{\varnothing} -2\ket{1234} 
\,,\quad
\mvS_1\ket{1234} =\ket{12}
+ \tfrac{1}{2}\ket{34}
\,,\\
&\mvS_1\ket{\epsilon_1,\epsilon_2,\epsilon_3,\epsilon_4} =
\tfrac{1}{2}\sum\limits_{\epsilon_3',\epsilon_4'}
(-1)^{\epsilon_3\epsilon_4'+\epsilon_4\epsilon_3'}
\ket{\epsilon_1,\epsilon_2,\epsilon_3',\epsilon_4'}
\,,\quad
\mvS_1\ket{ij} = \ket{ij}
\,,\quad (i=1,2,\quad j=3,4)\,.
\\
\end{split}
\ee
We can understand the action of $\mvS_1$ on the twisted sector states as follows.
First note that the orbifold CFT has a $\Z_2^4$ symmetry with elements that act as
$$
\Zivw(\varpi_1,\varpi_2,\varpi_3,\varpi_4)
\ket{\epsilon_1,\epsilon_2,\epsilon_3,\epsilon_4}
=(-1)^{\sum_{i=1}^4\varpi_i\epsilon_i}
\ket{\epsilon_1,\epsilon_2,\epsilon_3,\epsilon_4}\,.
$$
The CFT also has another $\Z_2^4$ symmetry corresponding to the discrete isometries \eqref{eqn:Ziv}. The ground states of the untwisted sector are invariant under these symmetries, but the ground states of the twisted sector are not:
$$
\Zivp(\varpi_1,\varpi_2,\varpi_3,\varpi_4)
\ket{\epsilon_1,\epsilon_2,\epsilon_3,\epsilon_4}
=\ket{\epsilon_1+\varpi_1,\epsilon_2+\varpi_2,\epsilon_3+\varpi_3,\epsilon_4+\varpi_4}.
$$
The operator $\mvS_1$ must convert discrete $\Z_2$ winding number in directions $3,4$ to discrete $\Z_2$ momentum in directions $4,3$, respectively, because $\mvS_1$ corresponds to T-duality. The symmetries $\Zivp(\varpi_1,\dots,\varpi_4)$ play the role of discrete momentum while $\Zivw(\varpi_1,\dots,\varpi_4)$ play the role of discrete winding number, thus
$$
\mvS_1^{-1}\Zivp(0,0,\varpi_3,\varpi_4)\mvS_1=
\Zivw(0,0,\varpi_4,\varpi_3).
$$
This explains the expression for $\mvS_1\ket{\epsilon_1,\dots,\epsilon_4}$ in \eqref{eqn:mvS1}. It is an eigenstate of the four $\Zivp(0,0,\varpi_3,\varpi_4)$'s with eigenvalues $(-1)^{\varpi_3\epsilon_4+\varpi_4\epsilon_3}$.

The transformation $\mvS_1$ corresponds to the T-duality element $\begin{pmatrix} 0 & 1 \\ -1 & 0 \\ \end{pmatrix}\in\SL(2,\Z)$ for the $T^2$ in directions $\xf_3,\xf_4,$ and we can also define $\mvT_1$ which corresponds to the element $\begin{pmatrix} 1 & 1 \\ 0 & 1 \\ \end{pmatrix}\in\SL(2,\Z)$.
 
\be\label{eqn:mvT1}
\begin{split}
&\mvT_1\ket{\varnothing} = \ket{\varnothing}
\,,\quad
\mvT_1\ket{12} =\ket{12}+ \ket{\varnothing}
\,,\quad
\mvT_1\ket{34} = \ket{34} 
\,,\quad
\mvT_1\ket{1234} =\ket{1234} - \tfrac{1}{2} \ket{34} 
\,,\\
&\mvT_1\ket{\epsilon_1,\epsilon_2,\epsilon_3,\epsilon_4} =
(-1)^{\epsilon_3\epsilon_4}
\ket{\epsilon_1,\epsilon_2,\epsilon_3,\epsilon_4}
\,,\quad
\mvT_1\ket{ij} = \ket{ij}
\,,\qquad (i=1,2,\quad j=3,4).
\\
\end{split}
\ee
Similarly to $\mvS_1,\mvT_1$ we define $\mvS_2,\mvT_2$ which act as T-duality on the $T^2$ in directions $\xf_1,\xf_2$. Thus, 
\be\label{eqn:mvS2}
\begin{split}
&\mvS_2\ket{\varnothing} = \ket{34}
\,,\quad
\mvS_2\ket{34} = -\ket{\varnothing}
\,,\quad
\mvS_2\ket{12} =2 \ket{\varnothing} -2\ket{1234}
\,,\quad
\mvS_2\ket{1234} =\ket{34}+ \tfrac{1}{2}\ket{12}
\,,\\
&\mvS_2\ket{\epsilon_1,\epsilon_2,\epsilon_3,\epsilon_4} =
\tfrac{1}{2}\sum\limits_{\epsilon_1',\epsilon_2'}
(-1)^{\epsilon_1\epsilon_2'+\epsilon_2\epsilon_1'}
\ket{\epsilon_1',\epsilon_2',\epsilon_3,\epsilon_4}
\,,\quad
\mvS_2\ket{ij} = \ket{ij}
\,,\quad (i=1,2,\quad j=3,4).
\\
\end{split}
\ee
and
\be\label{eqn:mvT2} 
\begin{split}
&\mvT_2\ket{\varnothing} = \ket{\varnothing}
\,,\quad
\mvT_2\ket{34} = \ket{34} + \ket{\varnothing}
\,,\quad
\mvT_2\ket{12} =\ket{12}
\,,\quad
\mvT_2\ket{1234} =\ket{1234} - \tfrac{1}{2} \ket{12} 
\,,\\
&\mvT_2\ket{\epsilon_1,\epsilon_2,\epsilon_3,\epsilon_4} =
(-1)^{\epsilon_1\epsilon_2}
\ket{\epsilon_1,\epsilon_2,\epsilon_3,\epsilon_4}
\,,\quad
\mvT_2\ket{ij} = \ket{ij}
\,,\qquad (i=1,2,\quad j=3,4).
\\
\end{split}
\ee


There is a map between the Fourier-Mukai vector space discussed in \secref{subsec:WittenIndexFM} and the vector space of ground states defined in \eqref{eqn:ketab}, 
\bear
&&x_0\mapsto\ket{\varnothing}
\,,\quad
x_4\mapsto
\ket{1234}-\tfrac{1}{4}\sum\limits_{\epsilon_1,\dots,\epsilon_4}
\ket{\epsilon_1,\epsilon_2,\epsilon_3,\epsilon_4}
\,,\nn\\ &&
[E(\epsilon_1,\dots,\epsilon_4)]\mapsto
\ket{\epsilon_1,\epsilon_2,\epsilon_3,\epsilon_4}-\tfrac{1}{2}\ket{\varnothing}
\,,\quad
[M(\mu,\nu)]\mapsto\ket{\mu\nu}
\,,\quad (1\le \mu<\nu\le 4)
\,,\nn\\ &&
\label{eqn:vtoRR}
\eear
We recall that \eqref{eqn:vtoRR} can be interpreted in terms of the phenomenon of ``fractional branes'' \cite{Diaconescu:1997br,Diaconescu:1999dt}. the Fourier-Mukai vector can be thought of as a vector of Ramond-Ramond charges carried by some D-brane configuration. Due to the coupling, for example, between the NSNS $2$-form field and the $1$-form RR-field on the world volume of a D$2$-brane, a D$2$-brane that wraps an exceptional divisor also carries the charge equivalent of half a D$0$-brane. 

Combining \eqref{eqn:mvS1} and \eqref{eqn:vtoRR} we get the action of $\mvS_1$ on the Fourier-Mukai vector $v$ via its action on the basis:
\be\label{eqn:mvS1FM}
\begin{split}
&\mvS_1(x_0) = [M(1,2)]
\,,\quad
\mvS_1([M(1,2)]) = -x_0
\,,\\
&\mvS_1([M(i,j)]) = [M(i,j)]
\,,\qquad (i=1,2,\quad j=3,4)
\,,\\
&\mvS_1([M(3,4)]) = -2x_4
-\tfrac{1}{2}\sum\limits_{\epsilon_1,\dots,\epsilon_4}
[E(\epsilon_1,\epsilon_2,\epsilon_3,\epsilon_4)]
-2x_0 
\,,\\
&\mvS_1([E(\epsilon_1,\epsilon_2,\epsilon_3,\epsilon_4)]) =
-\tfrac{1}{2}[M(1,2)]
+\tfrac{1}{2}\sum\limits_{\epsilon_3',\epsilon_4'}
(-1)^{\epsilon_3\epsilon_4'+\epsilon_4\epsilon_3'}
[E(\epsilon_1,\epsilon_2,\epsilon_3',\epsilon_4')]
+ \delta_{\epsilon_3 0}\delta_{\epsilon_4 0}x_0
\,,\\
&\mvS_1(x_4) = \tfrac{1}{2}[M(3,4)]+[M(1,2)]
-\tfrac{1}{2}\sum\limits_{\epsilon_1,\epsilon_2}
[E(\epsilon_1,\epsilon_2,0,0)] -x_0\,.
\\
\end{split}
\ee
The operation $\mvS_1$ is intuitively understood as T-duality on the $\xf_3,\xf_4$ directions. 

%

For the special complex structure $\tau=i$ there is an additional $\Z_4$ symmetry generated by a $(\pi/2)$-rotation of the $\xf_1-\xf_2$ torus around the origin. It acts as
\be\label{eqn:mvRot} 
\begin{split}
&\mvR_2\ket{\varnothing} = \ket{\varnothing}
\,,\quad
\mvR_2\ket{34} = \ket{34}
\,,\quad
\mvR_2\ket{12} =\ket{12}
\,,\quad
\mvR_2\ket{1234} =\ket{1234}
\,,\\
&\mvR_2\ket{\epsilon_1,\epsilon_2,\epsilon_3,\epsilon_4} =
\ket{\epsilon_2,\epsilon_1,\epsilon_3,\epsilon_4}
\,,\quad
\mvR_2\ket{1j} = \ket{2j}
\,,\quad
\mvR_2\ket{2j} = -\ket{1j}
\,,\qquad (j=3,4).
\\
\end{split}
\ee


We now compactify $\sigma^1$ on $S^1$ and insert a duality twist $\OpM$. We will start with the two cases with $\OpM=\mvS_1$ or $\OpM=\mvS_1\mvS_2$.
We wish to find the Hilbert space of ground states.
Using \eqref{eqn:ImvM} and the explicit expressions \eqref{eqn:mvS1}-\eqref{eqn:mvT1} and \eqref{eqn:mvS2}-\eqref{eqn:mvT2} we calculate the Witten indices:
\be\label{eqn:WI}
I(\mvS_1)=12,\qquad
I(\mvS_1\mvS_2)=8\,.
\ee
In \secref{subsubsec:mvS1}-\secref{subsubsec:mvS1mvS2}, we will describe the ground states directly by orbifolding the construction of \secref{sec:T2}.
We can also consider other duality twists.
Take for example the twists by $\mvS_1\mvT_1$ and $\mvS_1\mvT_1\mvS_2\mvT_2$ with calculated Witten indices:
\be\label{eqn:WIotherk}
I(\mvS_1\mvT_1)=10 ,\qquad
I(\mvS_1\mvT_1\mvS_2\mvT_2)=6.
\ee
These will be discussed in \secref{subsubsec:mvS1T1}-\secref{subsubsec:mvS1T1S2T2}.
Lastly, in \secref{subsubsec:mvS1R2} we will study a combination of rotation on the base $\efBase$ and duality on the fiber $\efFiber$ with the relevant Witten index
\be\label{eqn:WIS1R2}
I(\mvS_1\mvR_2)=4.
\ee

\subsubsection{The $\mvS_1$ twist}
\label{subsubsec:mvS1}

We need to account for a Witten index of $12$.
The twist $\mvS_1$ acts as T-duality on directions $\xf_3,\xf_4$.
Thus, we begin by adding 
$$
\frac{1}{2\pi}\int(\xf_3' d\xf_4-\xf_4' d\xf_3)\,,\qquad\text{with}\quad
\xf_I'\equiv\xf_I(\sigma^1=2\pi)\,,\quad
\xf_I\equiv\xf_I(\sigma^1=0)\,,
$$
to the action of the CFT.
The $\xf_1,\xf_2$ CFT is unchanged.
In the untwisted sector we set $\xf_I'=\xf_I$ for $I=1,2,3,4$.
At low-energy the $\xf_3-\xf_4$ sector of the CFT reduces to geometric quantization of $T^2$ with symplectic form $2d\xf_3\wedge d\xf_4$, resulting in a $2$-dimensional Hilbert space that corresponds to the space of level-$2$ $\Theta$-functions on $T^2$. All states of this sector are bosonic. The $\xf_1-\xf_2$ CFT reduces to supersymmetric quantum mechanics on $T^2$ with states corresponding to the cohomology $H^*(T^2,\R)$. We need to keep only the states that are invariant under the orbifold action $\xf_I\rightarrow -\xf_I$. This acts on level-$\lvk$ $\Theta$-functions as $\Theta_{\lvk,j}\rightarrow\Theta_{\lvk,\lvk-j}$ and so for $\lvk=2$ all $\Theta$-functions are invariant. On the base $\efBase$, the $\Z_2$-invariant states are those corresponding to the even cohomology, i.e., $1$ and $d\xf_1\wedge d\xf_2$, and the $\Z_2$-odd states are those that correspond to the odd cohomology, i.e., $d\xf_1$ and $d\xf_2$.
Now, the missing piece of information that we need in order to combine the states of the $\xf_3-\xf_4$ CFT with the states of the $\xf_1-\xf_2$ CFT, to form $\Z_2$-invariant states, is whether the single ground state of the fermionic sector of the $\xf_3-\xf_4$ directions is $\Z_2$-even or $\Z_2$-odd. If this ground state is $\Z_2$-even then we combine the $2$ states of the $\xf_3-\xf_4$ sector with the even dimensional cohomology states $1$ and $d\xf_1\wedge d\xf_2$. On the other hand, if the ground state is $\Z_2$-odd then we combine the $2$ states of the $\xf_3-\xf_4$ sector with the odd dimensional cohomology states $d\xf_1$ and $d\xf_2$. 
In any case, the untwisted sector produces $2\times 2=4$ states, but we need to know whether they contribute $+4$ or $-4$ to the Witten index.
An argument as in \secref{subsec:F} implies that the ground state of the fermionic $\psi^3,\psi^4,\bpsi^3,\bpsi^4$ system is actually $\Z_2$-odd and has an odd fermion number. Altogether, combining it with the $\Z_2$-odd and fermionic states $d\xf_3$ and $d\xf_4$ shows that the untwisted sector has $2\times 2=4$ states that contribute $+4$ to the Witten index.

In the twisted sector we have to set $\xf_I'=-\xf_I$ for $I=1,2,3,4$. At low-energy the $\xf_3-\xf_4$ sector reduces to geometric quantization of $T^2$ with symplectic form $-2d\xf_3\wedge d\xf_4$, again resulting in a $2$-dimensional Hilbert space. These states can still be identified with level-$2$ $\Theta$-functions provided we take the opposite complex structure where the holomorphic variable is $\bz'\equiv\xf_3+\xf_4\btau'$ instead of $z'\equiv\xf_3+\xf_4\tau$.
For the $\xf_1-\xf_2$ CFT the twisted boundary conditions leave us with only $4$ states that are localized at the fixed points of $\Z_2$, i.e., at 
\be\label{eqn:xfixed}
(\xf_1,\xf_2)=\text{one of $(0,0)$, $(0,\pi)$, $(\pi,0)$, $(\pi,\pi)$.}
\ee
They are all invariant under $\Z_2$ and the twisted sector therefore has $4\times 2=8$ states. They must have even fermion number so as to contribute $+8$ to the Witten index, so that altogether we get a total of $12$ bosonic states, in accordance with the Witten index calculated in \eqref{eqn:WI}. Note that the $4$ wave-functions of the states of the untwisted sector are spread across the base $\efBase$, while the $8$ wave-functions of the twisted sector are localized at the $4$ singular fibers, with $2$ states for each singular fiber.

\subsubsection{The $\mvS_1\mvS_2$ twist}
\label{subsubsec:mvS1mvS2}

The $\mvS_1\mvS_2$ twist acts as T-duality on both fiber and base.
The untwisted sector has $2\times 2=4$ states corresponding to geometric quantization of $T^2\times T^2$ with symplectic form
$$
\omega=2d\xf_1\wedge d\xf_2 + 2d\xf_3\wedge d\xf_4.
$$
This sector gives rise to $4$ states which can be identified with states of geometric quantization of $(T^2\times T^2)/\Z_2$ with symplectic form given by the same $\omega$ as above, which is Poincar\'e dual to $[M(1,2)]+[M(3,4)]$.

The twisted sector similarly has $2\times 2=4$  states corresponding to geometric quantization of $(T^2\times T^2)/\Z_2$ with symplectic form
$$
\omega'=-2d\xf_1\wedge d\xf_2 -2d\xf_3\wedge d\xf_4.
$$
This sector gives rise to $4$ states which can be identified with states of geometric quantization of $(T^2\times T^2)/\Z_2$ with symplectic form given by the same $\omega$ as above, which is Poincar\'e dual to $-[M(1,2)]-[M(3,4)]$.
Altogether in both sectors we get $8$ bosonic states, in accordance with the Witten index calculated in \eqref{eqn:WI}.


\subsubsection{The $\mvS_1\mvT_1$ twist}
\label{subsubsec:mvS1T1}

The twist $\mvS_1$ acts as T-duality on directions $\xf_3,\xf_4$ and the twist $\mvT_1$ acts as spectral flow on the $B$-field.
Thus, we add 
$$
\frac{1}{2\pi}\int(\xf_3' d\xf_4-\xf_4' d\xf_3 - \xf_3 d \xf_4 )\,,\qquad\text{with}\quad
\xf_I'\equiv\xf_I(\sigma^1=2\pi)\,,\quad
\xf_I\equiv\xf_I(\sigma^1=0)\,,
$$
to the action of the CFT.
The $\xf_1,\xf_2$ CFT is unchanged.
As before, we set $\xf_I'=\xf_I$ for $I=1,2,3,4$ in the untwisted sector.
Now at low-energy the $\xf_3-\xf_4$ sector of the CFT reduces to geometric quantization of $T^2$ with symplectic form $ d\xf_3\wedge d\xf_4$, with a $1$-dimensional Hilbert space that corresponds to the level-$1$ $\Theta$-function on $T^2$, which is invariant under the orbifold action $\xf_I\rightarrow -\xf_I$.  This state is bosonic. The $\xf_1-\xf_2$ CFT is unchanged from \ref{subsubsec:mvS1}, and since the ground state in the $\xf_3-\xf_4$ sector is again $\Z_2$-odd, we combine it with the odd cohomology states $d\xf_1$ and $d\xf_2$, and this contributes $+2$ to the Witten index.

In the twisted sector we have to set $\xf_I'=-\xf_I$ for $I=1,2,3,4$. At low-energy the $\xf_3-\xf_4$ sector reduces to geometric quantization of $T^2$ with symplectic form $-3d\xf_3\wedge d\xf_4$, with a $3$-dimensional Hilbert space. However, the Hilbert space  of level-$3$ $\Theta$-functions that are invariant under the orbifold action $\xf_I\rightarrow -\xf_I$ is $2$-dimensional;  $\Theta_{\bf{3} , 0 }$ and the linear combination $\Theta_{\bf{3}  ,1  }+\Theta_{\bf{3}  ,2} $ are invariant. These states are bosonic. 

The twisted sector of the $\xf_1-\xf_2$ CFT is again the same as in \ref{subsubsec:mvS1}, with $4$ states corresponding to the fixed points \eqref{eqn:xfixed}, so altogether the twisted sector has $4\times 2=8$ states. They must have even fermion number so as to contribute $+8$ to the Witten index, so that altogether we get a total of $10$ bosonic states, in accordance with the Witten index calculated in \eqref{eqn:WIotherk}.

\subsubsection{The $\mvS_1\mvT_1 \mvS_2\mvT_2$ twist}
\label{subsubsec:mvS1T1S2T2}

The $\mvS_1\mvT_1 \mvS_2\mvT_2$ twist acts as T-duality and spectral flow on both fiber and base.
The untwisted sector has $1\times 1=1$ state corresponding to geometric quantization of $T^2\times T^2$ with symplectic form
$$
\omega=d\xf_1\wedge d\xf_2 + d\xf_3\wedge d\xf_4.
$$
This state is $\Z_2$ even, and hence survives the orbifold projection of $(T^2\times T^2)/\Z_2$.
The twisted sector similarly has $2\times 2 + 1\times 1=5$  states which are the $\Z_2$-even states of the Hilbert space that we get by geometric quantization of $T^2\times T^2$ with symplectic form
$$
\omega'=-3d\xf_1\wedge d\xf_2 -3d\xf_3\wedge d\xf_4.
$$
The $2\times 2=4$ states come from the combination of $\Z_2$ invariant subspaces of the level-$3$ $\Theta$-functions on each $T^2$ (given by $\Theta_{\bf{3} , 0 }$ and $\Theta_{\bf{3}  ,1  }+\Theta_{\bf{3}  ,2} $ on each $T^2$).  The $1\times 1=1$ state comes from the combination of $1$-dimensional $\Z_2$ odd subspaces on each $T^2$ (given by $\Theta_{\bf{3}  ,1  }-\Theta_{\bf{3}  ,2} $ on each $T^2$).
Altogether, combining both sectors we get $6$ bosonic states, which is in accordance with the Witten index that we calculated in \eqref{eqn:WIotherk}.

\subsubsection{The $\mvS_1\mvR_2$ twist}
\label{subsubsec:mvS1R2}

The $\mvS_1$ twist acts on directions $\xf_3-\xf_4$ as in \eqref{subsubsec:mvS1}, while $\mvR_2$ acts as a rotation by $\pi/2$ of the $\xf_1-\xf_2$ torus, which is assumed to have complex structure $\tau=i.$ In the untwisted sector there are two ground states for the $\xf_3-\xf_4$ system, corresponding to $\Theta_{\lvk,j}$ for $\lvk=2$ ($j=0,1$), and there are two ground states of the $\xf_1-\xf_2$ system corresponding to the two $\mvR_2$ fixed points $(\xf_1,\xf_2)=(0,0)$ and $(\pi,\pi).$ In the twisted sector, we effectively add another $\Z_2$ twist, which is equivalent to changing the symplectic form from $2\xf_3\wedge\xf_4$ to $-2\xf_3\wedge\xf_4$ and at the same time replacing $\mvR_2$ with $\mvR_2^{-1}.$ The generator of the $\Z_2$ action acts trivially on the bosonic parts of the ground states of both the untwisted and twisted sectors, but the fermionic part of the twisted ground state has opposite $\Z_2$ charge relative to its untwisted counterpart. (This can be seen, for example, using the state-operator correspondence after bosonization of the fermions.) Thus, only one sector, say the untwisted one, survives the orbifold projection on $\Z_2$-invariant states. Altogether, we end up with four ground states, two for each fixed point $(\xf_1,\xf_2)=(0,0)$ and $(\pi,\pi).$

\section{Connection with geometric quantization}
\label{sec:ConnectGQ}

So far we have seen hints of a connection with geometric quantization of the target space.
In particular, for the simple case of a $T^2$ target space, the  relevant low-energy terms in the action reduce to the action of geometric quantization of $T^2$ at level $\lvk=2$ [see \eqref{eqn:I1T}], and thus we have identified the Hilbert space of ground states with the Hilbert space $\Hgq(T^2,\lvk)$ obtained by geometric quantization of $T^2$ with symplectic form $\frac{\lvk}{2\pi}d\xf_1\wedge d\xf_2$.
Geometric quantization defines a noncommutative geometry on $T^2$, and the latter made a well-known appearance in string theory in \cite{Connes:1997cr}-\cite{Seiberg:1999vs}. As explained in \cite{Seiberg:1999vs}, it appears when we take the limit of large $B$ field for a sigma-model with $T^2$ target space formulated on a worldsheet with boundary. In this limit the worldsheet action reduces to a 0+1D action of the form \eqref{eqn:I1LE} on the boundary (see also \cite{Cattaneo:1999fm}). Noncommutative geometry also makes an appearance on D-branes probing asymmetric orbifolds \cite{Blumenhagen:2000fp}. In our context, noncommutative geometry arises on a closed worldsheet, but with modified boundary conditions along a nontrivial cycle (i.e., the T-duality twist along the $S^1$). It would be interesting to understand whether there is a more general connection between duality-twists and geometric quantization, and we will explore this a little bit further in this section.

A connection with geometric quantization would imply that there is a natural way to construct operators that act on  $\Hgq(T^2,\lvk).$ The ring of operators can be constructed from the generators $\exp(i\xf_1)$ and $\exp(i\xf_2)$, and therefore we would like to understand how to construct matrix elements of these operators naturally from the $\sigma$-model.
Inserting $\exp(i\smX_j(\sigma_1))$ at some position $\sigma_1$ is not the right answer, because $\exp(i\xf_j)$ should be dimensionless, whereas the normal ordered operators  $:\exp(i\smX_j):$ have positive dimensions (which depend on the target space metric). As a matter of fact, even the operators $\exp(i\xf_j)$ in the 0+1D Landau problem [given by the Lagrangian $I_L$ of \eqref{eqn:LLag}] at low-energy do not flow directly to $\exp(i\xf_j)$ of the geometric quantization system --- there is a finite normalization constant in front.

A possible solution to this problem can be achieved as follows.
Let us complexify $\xf_j$ ($j=1,2$) by adding an imaginary part to define $\xyf_j\equiv\xf_j + i\yf_j$ with $\yf_j\in\R.$ We promote $\yf_j$ to 1+1D free bosonic fields and add fermionic superpartners. We now compactify as before with a T-duality twist on the $(\xf_1,\xf_2)$-directions, and insert a geometrical twist corresponding to a rotation by $(\pi/2)$ on the $(\yf_1,\yf_2)$-variables, namely:
$$
\yf_1(\sigma^1=0)=\yf_2(\sigma^1=2\pi)\,,\qquad
\yf_2(\sigma^1=0)=-\yf_1(\sigma^1=2\pi)\,,
$$
and similarly for the fermions, so as to preserve supersymmetry. This twist eliminates all zero modes, and leaves only one ground state in the $(\yf_1,\yf_2)$ system. In the combined $(\xf_1, \xf_2, \yf_1, \yf_2)$ system, a matrix element of products of operators $\exp(i\xyf_j)$ between ground states is independent of the insertion points $\sigma^1$ and reduces to the matrix element of the corresponding product of  $\exp(i\xf_j)$ operators in $\Hgq$. The contribution of the higher modes of the $\yf_j$'s cancels out the contribution of the higher modes of the $\xf_j$'s.
So we arrive at the identification of the low-energy limit (ground states) of a compactification of a $\sigma$-model with target space $T^2\times\R^2$ on $S^1$, with a combination of T-duality and geometrical twists, and geometric quantization of the $T^2$, which is the fixed point set of the geometrical component of the twist.

Let us now extend these observations to geometric quantization on $S^2$. We start with the orbifold $(T^2\times\R^2)/\Z_2$ and add a twisted boundary condition that acts as T-duality on $T^2$ and as rotation by $\pi/2$ on $\R^2.$ The wave-functions of the ground states fall off fast far away from the origin of $\R^2$. To count the number of ground states we recall the model of \secref{subsubsec:mvS1R2} which contained a similar twist, but with $\R^2$ replaced by $T^2$. The $\pi/2$ rotation in that model had two fixed points, but the behavior near each fixed point is the same as for the  $(T^2\times\R^2)/\Z_2$ model. Thus, the number of ground states of the  $(T^2\times\R^2)/\Z_2$ model is half that found in \secref{subsubsec:mvS1R2}, which is two ground states. These ground states should correspond to geometric quantization of $T^2/\Z_2$, which is the space at the origin of $\R^2$. The symplectic form is $\frac{\lvk}{4\pi^2}d\xf_1\wedge d\xf_2$ (for $\lvk=2$) whose integral over $T^2/\Z_2$ is $1$. The space $T^2/\Z_2$ is equivalent as a complex manifold to $\CP^1$, and so we expect that our system is equivalent to geometric quantization of $\CP^1\simeq S^2$ at level $1$. Indeed, the states of geometric quantization of $\CP^1$ at level $1$ are constructed as sections of the line bundle $\Obd(1)$, and the Hilbert space is indeed two-dimensional.
We can construct operators from $\Z_2$-invariant operators on $T^2$. Thus, we consider
\be\label{eqn:cosxy}
\Op_{n_1, n_2}\equiv
\cos(2\pi( n_1\xyf_1 + n_2 \xyf_2))\,.
\ee
Let us now describe how these operators flow in the IR to operators that arise by geometric quantization of $T^2/\Z_2\simeq\CP^1$. [That is to say, when projected to the Hilbert space of ground states the operators \eqref{eqn:cosxy} can be identified with operators acting on the Hilbert space of geometrically quantized $\CP^1$ at level $1$.]
First let us describe how to get sections of $\Obd(1)$ by $\Z_2$ projection of sections of a line bundle of degree $2$ on $T^2$.
We start with the $\Theta$-functions defined in \eqref{eqn:Thetafn}.
These functions satisfy
$$
\Theta_{j,\lvk}(u+1,\tau)=\Theta_{j,\lvk}(u,\tau)\,,\qquad
\Theta_{j,\lvk}(u+\tau,\tau)=e^{-2\pi i\lvk(u+\frac{1}{2}\tau)}\Theta_{j,\lvk}(u,\tau)\,,
$$
which makes them sections of a rank-$\lvk$ line-bundle over $T^2$.
This line-bundle is invariant under $\Z_2$, in the sense that if $f(u)$ is a section, i.e., satisfies
\be\label{eqn:fbc}
f(u+1)=f(u)\,,\qquad f(u+\tau)=e^{-2\pi i\lvk(u+\frac{1}{2}\tau)}f(u)\,,
\ee
then $g(u)\equiv f(-u)$ is also a section of the same bundle.
In particular, we have the relations
$$
\Theta_{j,\lvk}(-u,\tau)=\Theta_{j,\lvk-j}(u,\tau)\,.
$$
Since $f(u)$ and $f(-u)$ are sections of the same bundle, we can mod out by $\Z_2$ by imposing the condition $f(u)=f(-u).$ For even $\lvk$, an even holomorphic  function $f$ that satisfies the boundary conditions \eqref{eqn:fbc} projects to a holomorphic section of the line bundle $\Obd(\lvk/2)$ on $T^2/\Z_2\simeq\CP^1$. [For odd $\lvk$ there is a problem at the point $u=\tfrac{1}{2}(1+\tau)$, since as can be easily seen from \eqref{eqn:fbc}, the would-be section $f(\pm u)$ on $T^2/\Z_2$ would pick up a $(-)$ sign after traversing a small loop around  $\pm\tfrac{1}{2}(1+\tau)$.]
For $\lvk=2$ we have the stronger statement that all holomorphic sections are $\Z_2$-invariant. Thus, there is a natural map from the  ($\Z_2$-invariant) Hilbert space of ground states of geometric quantization of $T^2$ at level $\lvk=2$ to the Hilbert space of ground states of geometric quantization of $T^2/\Z_2\simeq\CP^1$ at level $\lvk=1$. The Hilbert space of ground states of a compactification of the $(T^2\times\R^2)/\Z_2$ theory with a T-duality and geometric rotation twist reduces to the Hilbert space of geometric quantization of the singular fiber $T^2/\Z_2$ at the origin of $\R^2$. The $\Z_2$ invariant operators \eqref{eqn:cosxy} reduce to the corresponding operators on the Hilbert space obtained by geometric quantization of $T^2/\Z_2$. Explicitly, if we denote by $\ket{j}$ the state that corresponds to $\Theta_{j,\lvk}$ then the projection on the ground states is calculated by geometric quantization and we have the ``clock and shift'' matrices
$$
e^{2\pi i\xf_1}\rightarrow\begin{pmatrix} 1 & 0 \\ 0 & -1 \\ \end{pmatrix}\,,\qquad
e^{2\pi i\xf_2}\rightarrow\begin{pmatrix} 0 & 1 \\ 1 & 0 \\ \end{pmatrix}\,,
$$
from which the general $\Op_{n_1, n_2}$ can easily be calculated and we find
$$
\cos(2\pi( n_1\xyf_1 + n_2\xyf_2))\rightarrow \begin{cases}
i^{n_1 n_2}\begin{pmatrix} 1 & 0 \\ 0 & (-1)^{n_1} \\ \end{pmatrix}\qquad \text{for even $n_2$}\,,\\
i^{n_1 n_2}\begin{pmatrix} 0 & 1\\ (-1)^{n_1} & 0\\ \end{pmatrix}\qquad \text{for odd $n_2$}\,.\\
\end{cases}
$$

\section{Discussion}
\label{sec:disc}

This paper was devoted to a few case-studies of the low-energy limit of a compactification of a quantum field theory on $S^1$ with boundary conditions twisted by a nonperturbative symmetry. The field theory was a 1+1D supersymmetric $\sigma$-model with $K_3$ target space, and the nonperturbative symmetry was an element of the duality group $O(20,4,\Z).$ The ``low-energy limit'' problem becomes the question of identifying the ground states.
We discussed several inequivalent elements of the duality group. Our main results refer to the element that can be interpreted as T-duality on the fiber, for an elliptically fibered $K_3$, where the K\"ahler modulus of the fiber is set to the self-dual value. We found that there must exist ground states with wave-functions that are localized at the points of the base (of the elliptic fibration) where the $T^2$ fiber degenerates. In addition to these, there are also states that have a spread-out wave-function. We showed that these states are in one-to-one correspondence with sections of certain holomorphic line-bundles over the $K_3$. We also supplemented the discussion with the analysis of other elements of $O(20,4,\Z)$ that we studied in the $T^4/\Z_2$ orbifold limit.

The association of a part of the Hilbert space of ground states with sections of a holomorphic line bundle $\Lbd$ over the $K_3$ target space suggests a relation with geometric quantization.
As was explained in \cite{Axelrod:1989xt}, the wave-functions of the Hilbert space obtained by geometric quantization of a K\"ahler manifold correspond to sections of a holomorphic line-bundle whose first Chern class is the class of the symplectic form (which is taken to be the K\"ahler class).
It would be interesting to formulate a more precise connection between the ground states of a compactification with a mirror-symmetry twist and geometric quantization. Such a connection should also contain a dictionary for mapping operators on the $\sigma$-model side to operators on the geometric quantization side. We have begun to explore such a connection in \secref{sec:ConnectGQ}, and we suggested that it might come about as follows. In a $\sigma$-model with a target space that is a $T^{2n}$ fibration over a $2n$-dimensional base, consider compactification on $S^1$ with a mirror-symmetry twist (realized locally as T-duality on the fiber) augmented by a suitable geometrical twist that is induced by an isometry of the base with isolated fixed points. For a suitably chosen geometrical twist, the Hilbert space of ground states might be identified with the states of geometric quantization of the fibers (which can be singular) over the fixed points of the geometrical twist. We hope to explore this idea further in upcoming work.

Our results could have applications to the study of S-duality of 3+1D \SUSY{4} Super-Yang-Mills (SYM) theory. In \cite{Ganor:2008hd,Ganor:2010md,Ganor:2012mu} a compactification of \SUSY{4} SYM on $S^1$ with boundary conditions given by a combination of S-duality and R-symmetry twists was studied, and a possible connection between the low-energy limit and Chern-Simons theory was suggested, but is not very well understood or established. (See also \cite{Terashima:2011qi} for another interesting approach.)
The connection with our present setting appears if we compactify the 3+1D theory on a Riemann surface $\Sigma_g$ (of genus $g$), together with a suitable topological twist. If $\Sigma_g$ shrinks first, the SYM theory reduces to a certain 1+1D supersymmetric $\sigma$-model \cite{Harvey:1995tg,Bershadsky:1995vm} (and see \cite{Kapustin:2006pk} for a thorough review and new perspective). The target space is the Hitchin moduli space $\MHitchin$ \cite{Hitchin:1986vp} associated with the gauge group $G$ and Riemann surface $\Sigma_g$, and S-duality reduces to mirror-symmetry of $\MHitchin$. For $g>1$ the Hitchin space has (complex) dimension $2(g-1)\dim G$ (where $\dim G$ is the dimension of the gauge group) and can be represented as a fibration \cite{Hitchin:1986vp} of $T^{2(g-1)\dim G}$ over a $(g-1)\dim G$ (complex) dimensional base $\efBase$ (and the fibers are allowed to degenerate over codimension-$1$ submanifolds of $\efBase$). It is therefore interesting to study the ground states of a compactification of this $\sigma$-model on $S^1$ with a mirror-symmetry twist. We expect that there are two kinds of ground states: (i) those with wave-functions that can be expanded in terms of (holomorphic) $\Theta$-functions of the fibers, with coefficients that are forms on the base (and it should be possible to map these to the cohomology with coefficients in a suitable line-bundle over $\MHitchin$); and (ii) states with wave-functions that are localized at the degenerate fibers. We hope to explore this subject and its connection to S-duality of SYM in more detail in future work.


\acknowledgments
We wish to thank Bruno Zumino for helpful discussions.
This work was supported in part by the Berkeley Center of Theoretical Physics,
and in part by the U.S. National Science Foundation under grant PHY-10-02399.

%
%
%


\bibliographystyle{my-h-elsevier}

\end{document}